\documentclass[PAPER, atlasdraft=false, USenglish, cernpreprint=false, cernpreprint, texmf, orcidlogo]{atlasdoc}
\usepackage[subfigure]{atlaspackage}
\usepackage{atlasbiblatex}

\usepackage[jetetmiss]{atlasphysics}
\graphicspath{{logos/}{figures/}}

\usepackage{ANA-STDM-2022-05-PAPER-defs}

\AtlasTitle{Measurement of jet track functions in $pp$ collisions at $\sqrt{s}=13$~\TeV with the ATLAS detector}

\AtlasAbstract{%
Measurements of jet substructure are key to probing the energy frontier at colliders, and many of them use track-based observables which take advantage of the angular precision of tracking detectors.
Theoretical calculations of track-based observables require `track functions', which characterize the transverse momentum fraction $r_q$ carried by charged hadrons from a fragmenting quark or gluon.
This letter presents a direct measurement of $r_q$ distributions in dijet events from the $140$~\ifb of proton--proton collisions at $\sqrt{s}=13$~\TeV recorded with the ATLAS detector.
The data are corrected for detector effects using machine-learning methods.
The scale evolution of the moments of the $r_q$ distribution is sensitive to non-linear renormalization group evolution equations of QCD, and is compared with analytic predictions.
When incorporated into future theoretical calculations, these results will enable a precision program of theory-data comparison for track-based jet substructure observables.

}

\AtlasRefCode{STDM-2022-05}
\PreprintIdNumber{CERN-EP-2024-333}

\AtlasJournalRef{\PLB 868 (2025) 139680}
\AtlasDOI{https://doi.org/10.1016/j.physletb.2025.139680}

\hypersetup{pdftitle={ATLAS document},pdfauthor={The ATLAS Collaboration}}

\begin{document}

\maketitle

\newcommand{\AtlasCoordFootnote}{%
ATLAS uses a right-handed coordinate system with its origin at the nominal interaction point (IP)
in the centre of the detector and the \(z\)-axis along the beam pipe.
The \(x\)-axis points from the IP to the centre of the LHC ring,
and the \(y\)-axis points upwards.
Cylindrical coordinates \((r,\phi)\) are used in the transverse plane,
\(\phi\) being the azimuthal angle around the \(z\)-axis.
The pseudorapidity is defined in terms of the polar angle \(\theta\) as \(\eta = -\ln \tan(\theta/2)\).
Angular distance is measured in units of \(\Delta R \equiv \sqrt{(\Delta\eta)^{2} + (\Delta\phi)^{2}}\).}

\section{Introduction}

Jets are collimated streams of hadrons resulting from the fragmentation of high energy quarks and gluons.
The wide range of energy scales probed during jet formation provides a unique experimental window into the dynamics of Quantum Chromodynamics (QCD) in both perturbative and non-perturbative regimes of the strong coupling.
The experimental study of jets has greatly expanded in the last decade with the ability to measure correlations between multiple particles produced in the same identified jet, a field referred to as jet substructure~\cite{Larkoski:2017jix,Kogler:2018hem}.
Jet substructure has provided numerous new ways to study the dynamics of the Standard Model and to search for new phenomena~\cite{Kogler:2018hem}.

The theoretical understanding of jets is based on the combination of perturbation theory and non-perturbative functions characterizing the hadronization process, which are universal in the context of factorization.
The most well-known examples of such a universal non-perturbative object are fragmentation functions~\cite{Collins:1981uw}, which describe the energy distribution of single hadrons arising from the fragmentation of quarks and gluons.
These functions cannot currently be calculated from first principles, and therefore must be determined by experiment.
The broad program of fragmentation-function measurements~\cite{STDM-2017-16} has been extremely successful, enabling the theoretical understanding of a wide variety of collider physics processes.
However, fragmentation functions are single-hadron observables, and therefore do not incorporate the correlations in the hadronization process necessary for the understanding of modern jet substructure observables.

Among the many functions characterizing the hadronization process, those describing the fragmentation into electrically charged hadrons, which are identified as tracks in the detector, are particularly important.
Tracking detectors have better angular granularity, enabling measurements at small angular scales within jets.
Track-based measurements have been key to enabling a precision jet substructure program at the Large Hadron Collider (LHC) by the ATLAS~\cite{STDM-2010-14,STDM-2014-17,STDM-2017-16,STDM-2017-33,STDM-2018-57,STDM-2020-31}, CMS~\cite{CMS-SMP-20-010,CMS-SMP-22-007}, ALICE~\cite{ALargeIonColliderExperiment:2021mqf,ALICE:2019ykw,ALICE:2021njq,ALICE:2021aqk} and LHCb~\cite{LHCb:2017llq,LHCb:2019qoc} experiments.

The interpretation of track-based jet substructure measurements requires an understanding of the correlated fragmentation of quarks and gluons into charged hadrons.
This is described by universal non-perturbative functions called track functions~\cite{Chang:2013rca,Chang:2013iba}, which describe the energy distribution of charged hadrons arising from fragmenting quarks or gluons.
These functions are multi-hadron observables.
Experimentally, track functions can be determined~\cite{Lee:2023xzv} from a measurement of the the transverse-momentum (\pt) fraction in tracks in high-energy jets denoted by $r_q=\tf$, where \ptall~ is the \pt of the jet and \ptch~ is the \pt of the sum of charged particles within the jet.
The first moment of this distribution, $\langle r_q \rangle\sim 2/3$ due to approximate isospin symmetry, is described by standard fragmentation functions, and is a fundamental property of the strong force that has been utilized by a multitude of experiments for calibration and precision studies~\cite{JETM-2018-02}.
The higher moments of the distribution encode interesting correlations in the hadronization process, and have not previously been measured.
In addition to their explicit values, which are an important ingredient for precision jet substructure calculations using tracks~\cite{Jaarsma:2023ell,Chen:2020vvp}, their scale evolution provides a test of QCD beyond the DGLAP paradigm~\cite{Gribov:1972ri,Dokshitzer:1977sg,Altarelli:1977zs}, yielding interesting insights into the non-linear renormalization group evolution of correlations in the hadronization process~\cite{Chen:2022pdu,Chen:2022muj,Jaarsma:2022kdd,Li:2021zcf}.

This paper presents a measurement of the energy distribution of tracks in identified high-\pt jets, using the full Run 2 dataset of proton--proton ($pp$) collisions at $\sqrt{s}=13$~\TeV recorded with the ATLAS detector at the LHC, corresponding to an integrated luminosity of $140$~\ifb.
The measurement is performed differentially in \rcharged and in different regions of jet \pt and for different rapidity-ordering of a dijet system (\emph{i.e.,} for more-central \emph{vs.} more-forward jets). These distributions are used to extract the moments of \rcharged as a function of the jet \pt.
The data are corrected for detector effects using Iterative Bayesian Unfolding (IBU)~\cite{Lucy:1974yx,Richardson:72,DAgostini:1995} and a machine learning-based method called OmniFold~\cite{1911.09107,Andreassen:2021zzk} is employed as a novel data-driven correction for binning artifacts in the moment extractions.

\section{ATLAS Detector}

The ATLAS experiment~\cite{PERF-2007-01} at the LHC is a multipurpose particle detector with a forward--backward symmetric cylindrical geometry and a near \(4\pi\) coverage in solid angle.\footnote{\AtlasCoordFootnote}
It consists of an inner tracking detector surrounded by a thin superconducting solenoid providing a \qty{2}{\tesla} axial magnetic field, electromagnetic and hadronic calorimeters, and a muon spectrometer.
The inner tracking detector covers the pseudorapidity range \(|\eta| < 2.5\). It consists of silicon pixel, silicon microstrip, and transition radiation tracking detectors.
Lead/liquid-argon (LAr) sampling calorimeters provide electromagnetic (EM) energy measurements with high granularity within the region \(|\eta|< 3.2\). A steel/scintillator-tile hadronic calorimeter covers the central pseudorapidity range (\(|\eta| < 1.7\)). The endcap and forward regions are instrumented with LAr calorimeters for EM and hadronic energy measurements up to \(|\eta| = 4.9\).
The muon spectrometer surrounds the calorimeters and is based on three large superconducting air-core toroidal magnets with eight coils each. The field integral of the toroids ranges between \num{2.0} and \qty{6.0}{\tesla\metre} across most of the detector. The muon spectrometer includes a system of precision tracking chambers up to \(|\eta| = 2.7\) and fast detectors for triggering up to \(|\eta| = 2.4\).
The luminosity is measured mainly by the LUCID--2~\cite{LUCID2} detector, which is located close to the beampipe.
A two-level trigger system is used to select events~\cite{TRIG-2016-01}. The first-level trigger is implemented in hardware and uses a subset of the detector information to accept events at a rate close to \qty{100}{\kHz}. This is followed by a software-based trigger that reduces the accepted rate of complete events to \qty{1.25}{\kHz} on average depending on the data-taking conditions.
A software suite~\cite{ATLAS:2024vdo} is used in data simulation, in the reconstruction and analysis of real and simulated data, in detector operations, and in the trigger and data acquisition systems of the experiment.

\section{Simulated event samples}

Samples of Monte Carlo (MC) simulated dijet events are used to perform the unfolding and compare with the corrected data.
\PYTHIA[8.230]~\cite{Sjostrand:2014zea,Sjostrand:2006za} is used as the nominal MC generator for this analysis, and is also referred to here as the `nominal' simulation.
Samples of $\TwoToTwo$ dijet events were simulated at leading-order (LO) in QCD using the A14 tune~\cite{ATL-PHYS-PUB-2014-021}, the Lund string hadronisation model and the \NNPDF[2.3lo]~\cite{Ball:2012cx} parton distribution function (PDF) set.
The \PYTHIA parton shower (PS) algorithm uses a dipole-style \pt-ordered evolution, and its renormalisation and factorisation scales were set to the geometric mean of the squared transverse masses of the outgoing particles.
\EVTGEN~\cite{Lange:2001uf} was used to model decays of heavy-flavour hadrons.

Two sets of \SHERPA[2.2.5]~\cite{Gleisberg:2008ta} dijet events were simulated with the default AHADIC cluster hadronisation model~\cite{Webber:1983if} or with the \SHERPA interface to the Lund string hadronisation model as implemented in \PYTHIA[6.4], and its decay tables.
These samples include LO matrix element calculations for \TwoToTwo processes, and use the \SHERPA parton shower algorithm based on Catani--Seymour dipole subtraction~\cite{Schumann:2007mg}.
The \CT[14nnlo] next-to-next-to-leading-order (NNLO) PDF~\cite{Dulat:2015mca} set is used for matrix element calculations and \CT[10] is used for multi-parton interactions (MPI)~\cite{Lai:2010vv}.

A sample of \HERWIG[7.1.3]~\cite{Bahr:2008pv,Bellm:2015jjp,Bellm:2017bvx} multijet events was generated with the \MMHT[nlo] PDF set~\cite{Harland-Lang:2014zoa}, default cluster hadronisation model and the default angle-ordered PS.
This sample models \TwoToTwo~matrix elements with NLO accuracy and~\TwoToThree~matrix elements with LO accuracy.
The parton shower was matched to the matrix element calculation using the \MCatNLO matching scheme~\cite{Lonnblad:2001iq,Lonnblad:2011xx}, and the \pt of the leading jet is taken as the renormalisation scale.

All generated events used in the unfolding and its associated uncertainties were passed through a full detector simulation~\cite{SOFT-2010-01} based on \GEANT~\cite{Agostinelli:2002hh} and overlaid with simulated minimum-bias interactions generated using \PYTHIA[8]~with the A3 tune~\cite{ATL-PHYS-PUB-2016-017} and \NNPDF[2.3lo] PDF set~\cite{Ball:2012cx} to represent the effect of multiple $pp$ interactions in the same and neighboring bunch crossings (`pile-up').
The distribution of the average number of pile-up interactions in simulation was reweighted during data analysis to match that observed in \RunTwo data.
Additional details of the MC samples used in this measurement may be found in Ref.~\cite{ATL-PHYS-PUB-2019-017}.

\section{Event reconstruction and selection}

Detector-level inner-detector tracks are required to have $\pt > 500~\MeV$, to be associated with the primary vertex with the largest sum of track $\pt^2$ in the event, and to satisfy `loose' quality criteria~\cite{ATL-PHYS-PUB-2015-026}.
At particle-level, charged hadrons (and leptons produced through secondary decays) are required to have $\pt > 500~\MeV$.
Tracks (charged hadrons) within $\DeltaR=0.4$ of the cores of selected detector-level (particle-level) jets are clustered using the \akt algorithm~\cite{Cacciari:2008gp} as implemented in \Fastjet~\cite{Fastjet} in order to obtain the jet \pt originating only from charged particles, \emph{i.e.} \ptch.

Detector-level jets are reconstructed from particle-flow objects~\cite{PERF-2015-09}
using the \akt algorithm with radius parameter $R=0.4$~\cite{Cacciari:2008gp,Fastjet}, which combine measurements from the ATLAS inner detector and calorimeter systems to improve the jet energy resolution (JER) and the jet reconstruction efficiency, especially for low-\pt jets.
The jet energy scale (JES) is calibrated so that, on average, the detector-level jet energy is the same as that of the corresponding particle-level jets~\cite{JETM-2018-05}.
These jets are `cleaned' to remove those originating from detector noise, cosmic rays and beam-induced processes~\cite{ATLAS-CONF-2015-029}.
Merging pixel clusters within the dense core of jets at detector-level may degrade the track parameter resolution and decrease the track reconstruction efficiency.
These effects are mitigated with the use of stacked neural networks that assist with ambiguity-solving during track reconstruction by splitting clusters of pixel charges~\cite{PERF-2015-08, ATL-PHYS-PUB-2015-006, ATL-PHYS-PUB-2017-016, ATL-PHYS-PUB-2015-052,ATL-PHYS-PUB-2015-044}.
Particle-level jets are reconstructed in MC generated events without detector simulation.
All particles with a laboratory-frame lifetime $\tau$ such that $c\tau > 10$~mm are used, except those particles that are expected to leave no or negligible energy depositions in the calorimeter (\emph{i.e.}\ neutrinos or muons).

Events in data are selected using single-jet triggers~\cite{TRIG-2016-01,TRIG-2012-01} and the highest two $\pt$ jets are analyzed.
The leading and subleading jets are selected for the measurement at both detector- and particle-level, and are required to satisfy $\pt^\text{leading} >240~\GeV$ and $\pt^\text{leading} < 1.5\times \pt^\text{subleading}$.
This balance requirement simplifies the interpretation of the final state in terms of a \TwoToTwo scattering process.
Both jets must be completely within the inner detector acceptance ($|\eta_{\text{jet}}|<2.1$).

\section{Unfolding procedure}

The selected data are unfolded to correct for detector bias, resolution, and acceptance effects by applying IBU~\cite{DAgostini:1995} with two iterations implemented in RooUnfold~\cite{Adye:2011gm}.
The MC generator used to unfold the data is \PYTHIA[8.230], and the total number of iterations is chosen to minimize the total uncertainty.
The unfolding procedure corrects the \rcharged constructed from detector-level objects to particle level, where the same kinematic requirements are imposed on particle-level jets as on detector-level jets.

Particle-level jets not passing the detector-level kinematic requirements, and detector-level jets not passing the particle-level requirements, are accounted for with efficiency and purity corrections, respectively.
Purity corrections are applied before the regularized inversion of the response matrix, and efficiency corrections are applied afterwards.
One additional low jet \pt bin is included in the unfolding matrix so migrations into that bin take the place of (much of) the inefficiency.
The jet \pt, \rcharged, and $\eta$ region are unfolded simultaneously, where the $\eta$ region of the two selected jets (denoted `Forward $\eta$ bin' and `Central $\eta$ bin', respectively).
The binning of the jet \pt and of the \rcharged distribution is chosen such that the migrations in the unfolding matrix are typically less than 60\%.
The migrations are smallest at low \rcharged and low jet \pt, with larger migrations in the higher \pt bins where the track \pt resolution degrades.
For each jet \pt bin, the unfolded distribution is normalized to the number of jets in that bin.

\section{Uncertainties}

Statistical and  systematic uncertainties originating from several different sources are considered in this measurement.
Different sources of uncertainty are always treated independently.

The statistical uncertainties due to the finite statistics of the data and simulated event samples are evaluated from 100 pseudo-experiments using the bootstrap method~\cite{jackknife}.
The data and simulation statistical uncertainties are typically subdominant.
The statistical uncertainty from data is larger than that originating from MC samples throughout the measurement.

Theoretical uncertainties arise mainly from the accuracy of jet fragmentation modeling, as these discrepancies can modify the unfolding efficiency/purity corrections and the response matrix.
The unfolding is repeated using the \PYTHIA[8] prior and a response matrix constructed using the \SHERPA[2.2.5] AHADIC sample instead of the nominal \PYTHIA[8] one.
The difference between this result and the nominal one is taken as a systematic uncertainty.
This uncertainty is typically $\sim5\%$ in the tails of \rcharged, and $\sim2\%$ in the bulk of the \rcharged distribution, and is one of the dominant sources of uncertainty for large \rcharged values in the higher bins of jet \pt (beginning at $\pt>600~\GeV$).

Experimental systematic uncertainties are evaluated by propagating variations from different sources through the unfolding procedure to find the difference between the varied and nominal results.
Uncertainties on the jet energy scale and resolution are determined using a mixture of simulation-based and \insitu techniques~\cite{JETM-2018-05}.
Uncertainties due to differences between the gluon-initiated jet energy response of different MC generator setups have also been reduced (‘jet flavor response’ in Ref.~\cite{JETM-2018-05}) by performing more granular comparisons of the effect of different parton shower and hadronization models on the jet response as documented in Ref.~\cite{STDM-2020-04}.
These uncertainties cause the migration of jets into or out of the fiducial acceptance and modify the denominator of the \rcharged observable, typically accounting for uncertainties around 2\% in total, but reaching up to 3.5\% in some bins.
Uncertainties related to the reconstruction of isolated tracks and tracks within dense environments are considered by modifying the measured \pt of individual tracks or removing them from the event~\cite{PERF-2015-08,PERF-2015-07}.
The tracking uncertainty is the largest source of uncertainty for \rcharged for all but the highest jet \pt bin.
The dominant uncertainty in this category is related to the inclusive tracking efficiency.
This uncertainty is largest for small values of \rcharged and is up to 8\% in the more central $\eta$ bin and 12\% in the more forward $\eta$ bin.
Potential biases arising due to differences between disabled modules of the Tile calorimeter during certain data-taking periods between the data and simulation are evaluated.
This is done by repeating the measurement with a veto on jets directed at any disabled modules in both data and simulation, and comparing this result to the nominal result.
These uncertainties are typically between $2-3\%$, with the largest impact at lower values of \rcharged.
An uncertainty is added to cover unfolding regularization biases, which is determined by unfolding the detector-level \PYTHIA distribution after it is reweighted based on a comparison of the corresponding simulated detector-level distribution and the detector-level data prior~\cite{Malaescu:2009dm}.
This uncertainty is largest for small values of \rcharged and increases with the jet \pt.
This uncertainty is typically subdominant, but in the highest jet \pt bin, it is the dominant source of uncertainty, reaching up to 12\%.
Uncertainties related to the overlaid pileup events amount to an error of less than 1\% throughout the measurement.

\section{Results}

The measured cross-section as a function of $r_q$ is shown for a representative bin of jet \pt from 600--800~\GeV, for the central and forward regions in Figure~\ref{fig:xsec}.
These measured distributions are compared to predictions from several MC generators, which typically agree with the data within uncertainties.
The different MC predictions show the same trends, tending to underestimate the cross-section at low values of \rcharged and overestimate the cross-section at high values.
These trends are consistent across all measured \pt bins.

The \rcharged distributions are used to extract the moments, $\text{E}[X^n]$, as a function of the jet \pt.
These moments are shown in Figure~\ref{fig:moment}, and are found to evolve slowly with the jet \pt.
The binning of \rcharged introduces artifacts in the moment extraction, and these are corrected for using a data-driven method based on the OmniFold algorithm~\cite{1911.09107,Andreassen:2021zzk}.
In this method, the unbinned data (charge sum and \pt for each jet - for features per event) are unfolded using OmniFold, and multiplicative correction factors are determined by comparing the moments with and without the binning.
This improves upon the MC-based binning corrections that have been applied in previous measurements~\cite{STDM-2014-17,STDM-2017-16} by directly utilizing the data for the correction.
The correction factors are applied to all moments extracted from the \rcharged distribution accordingly, after each systematic variation, where the magnitudes of the correction factors range from 0.92 to 1.
This approach is conservative, as it preserves the absolute uncertainty of each systematic variation on the uncorrected distribution following the binning correction.

\begin{figure}[t!] %
\centering
\subfigure[]{\includegraphics[width=0.49\textwidth]{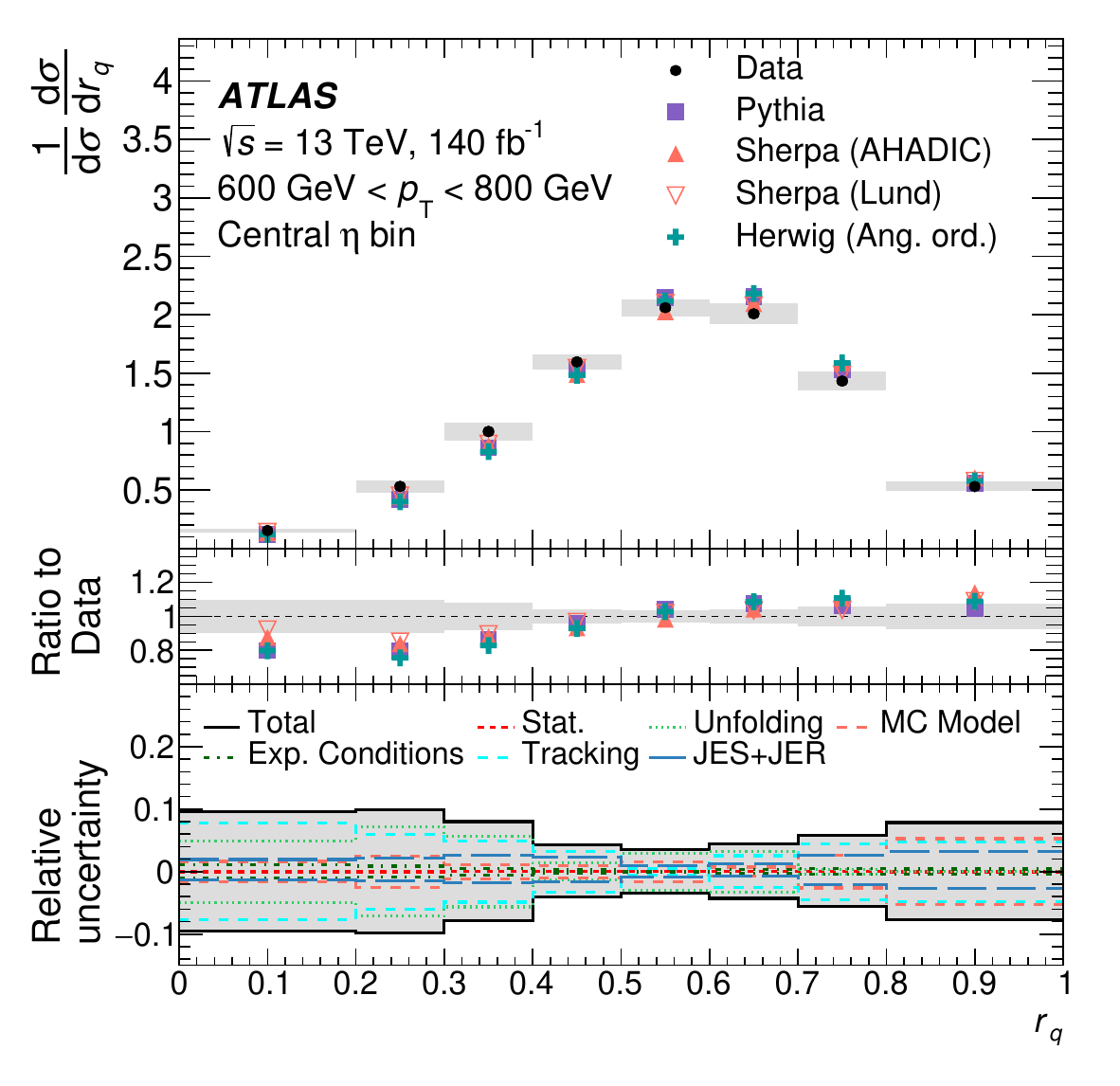}\label{fig:xsec:a}}
\subfigure[]{\includegraphics[width=0.49\textwidth]{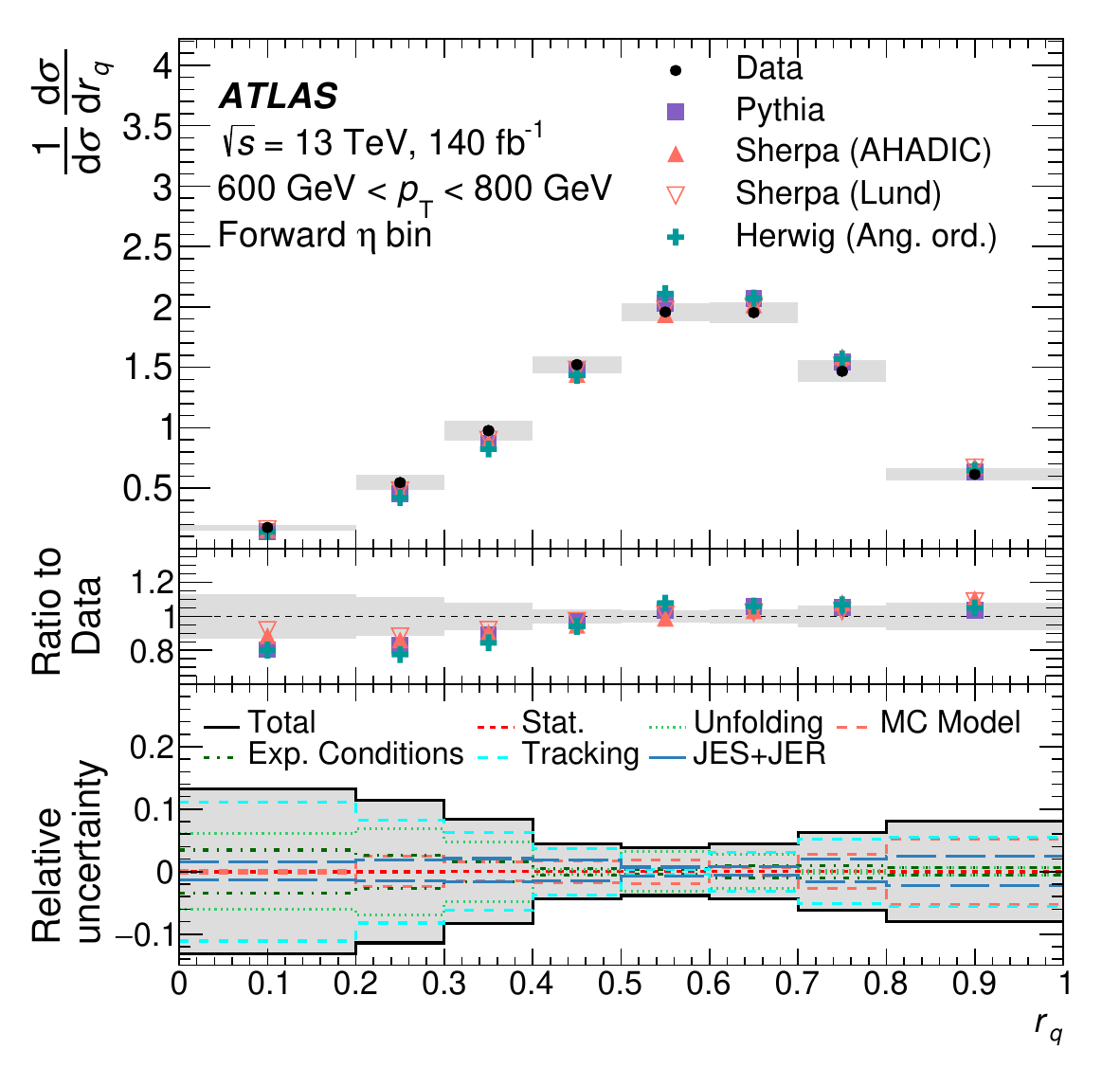}\label{fig:xsec:b}}
\caption{The unfolded central (a) and forward (b) normalized differential cross-sections as a function of \rcharged for data compared to predictions from several MC generators. The grey uncertainty band shows the combined statistical and systematic uncertainties on the measurement.
The bottom panel shows the breakdown of the uncertainties grouped by their sources.
In particular, the `Exp. Conditions' group includes uncertainties related to the overlaid pileup events and potential biases due to differences between disabled modules of the Tile calorimeter.
}
\label{fig:xsec}
\end{figure}

\begin{figure}[t!] %
\centering
\subfigure[]{\includegraphics[width=0.49\textwidth]{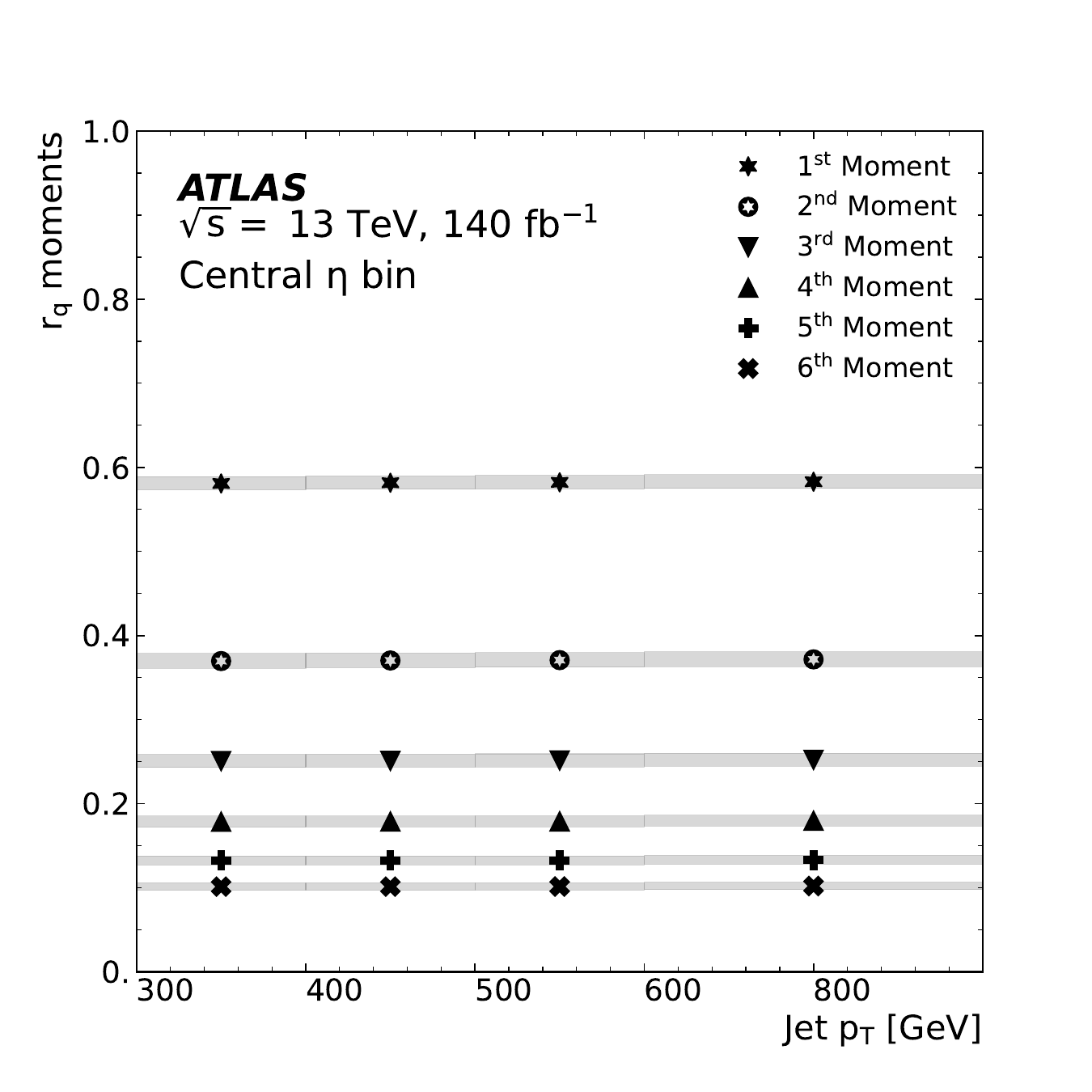}}
\subfigure[]{\includegraphics[width=0.49\textwidth]{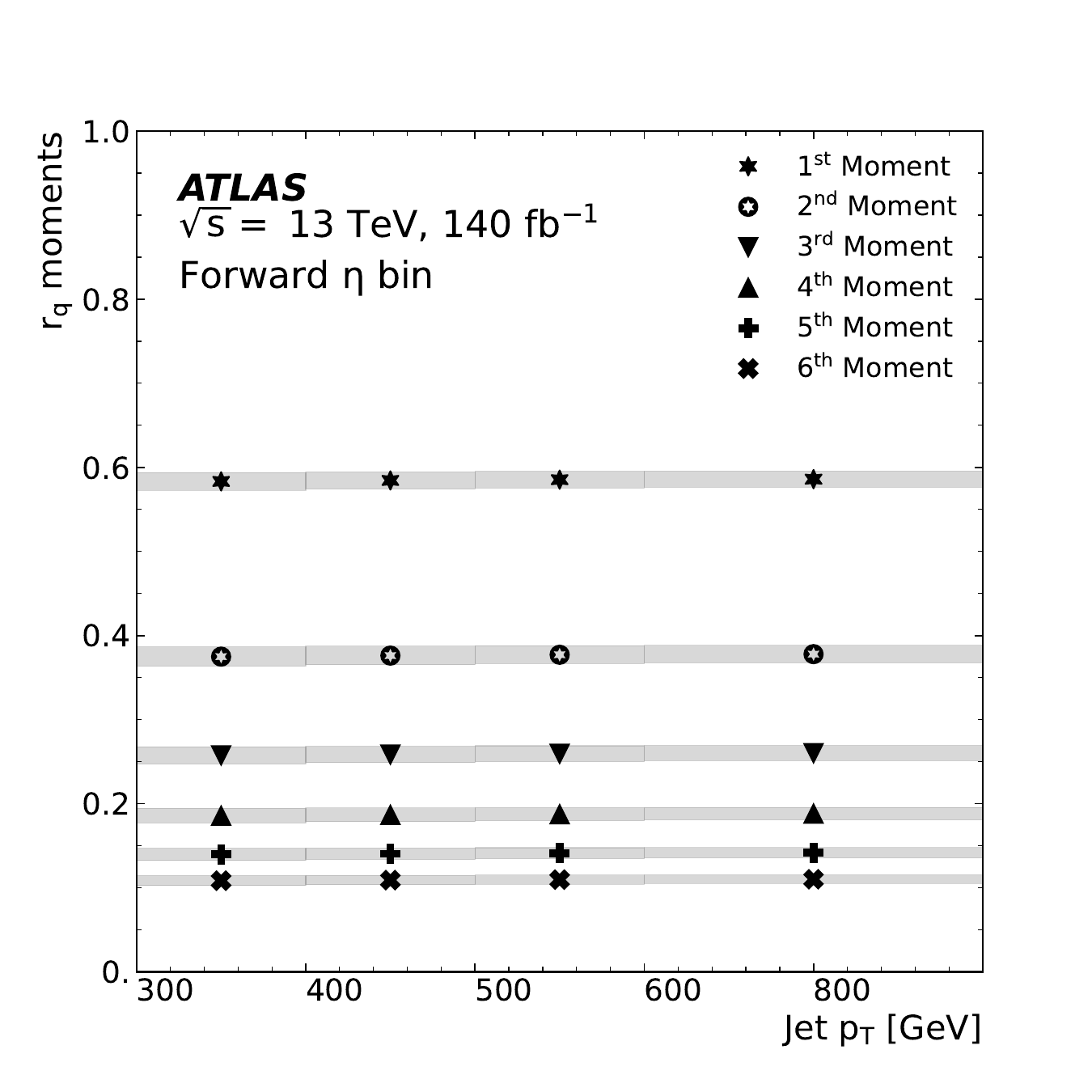}}
\caption{The first six moments of the unfolded (a) central and (b) forward \rcharged distributions in jet \pt bins. The grey uncertainty band shows the combined statistical and systematic uncertainties on the measurement.
}
\label{fig:moment}
\end{figure}

\begin{figure}[t!] %
\centering
\subfigure[]{\includegraphics[width=0.49\textwidth]{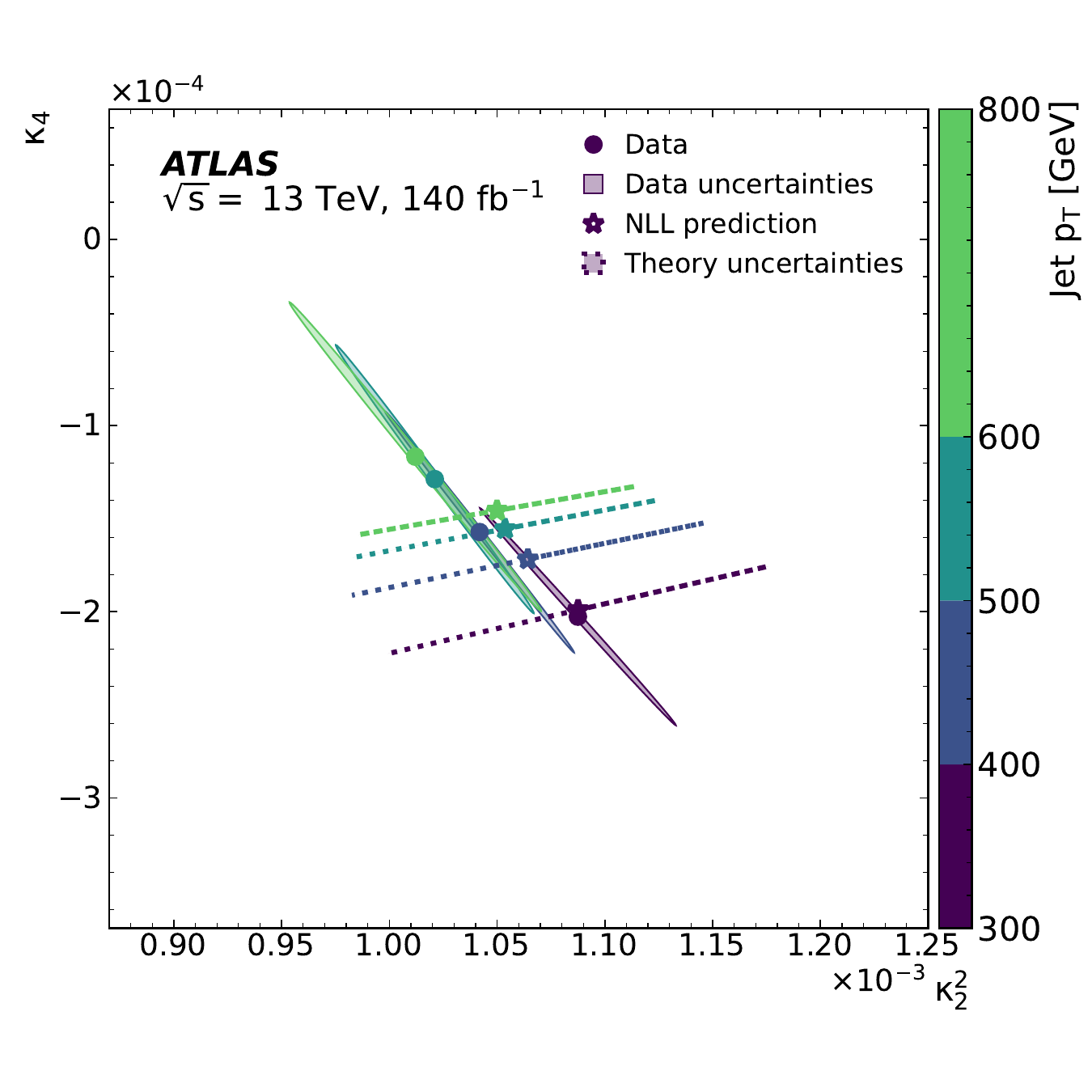}\label{fig:flow:a}}
\subfigure[]{\includegraphics[width=0.49\textwidth]{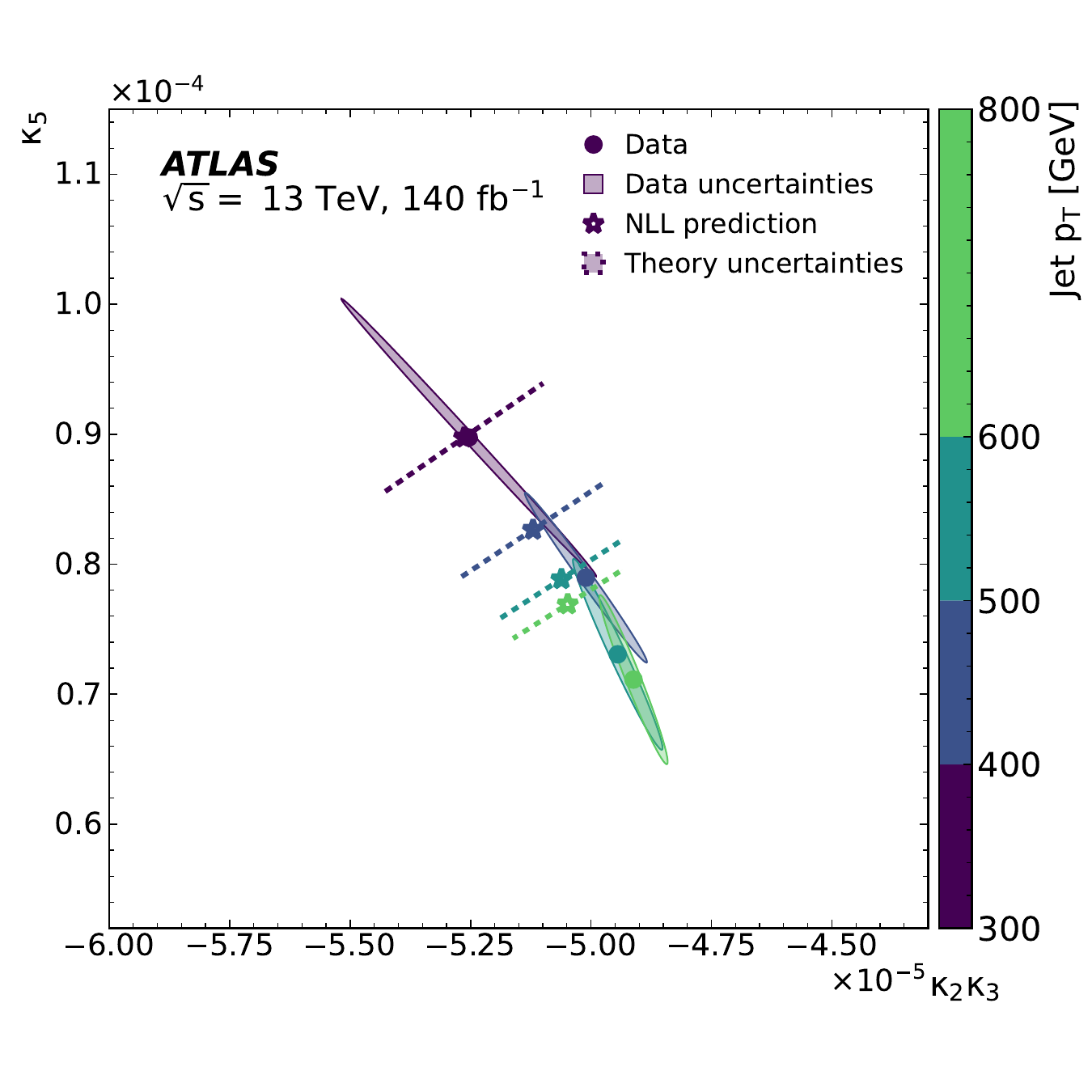}\label{fig:flow:b}}
\subfigure[]{\includegraphics[width=0.49\textwidth]{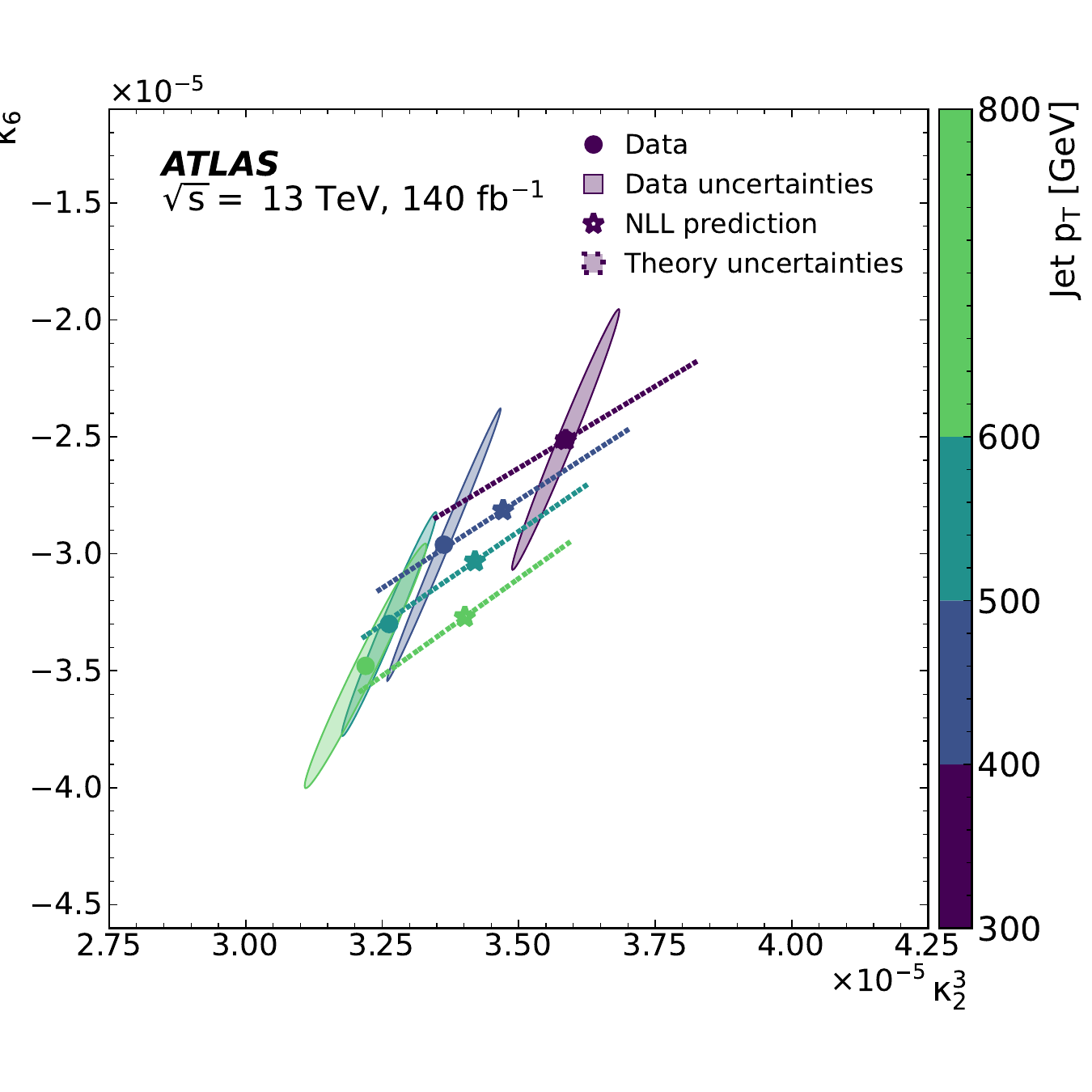}\label{fig:flow:c}}
\subfigure[]{\includegraphics[width=0.49\textwidth]{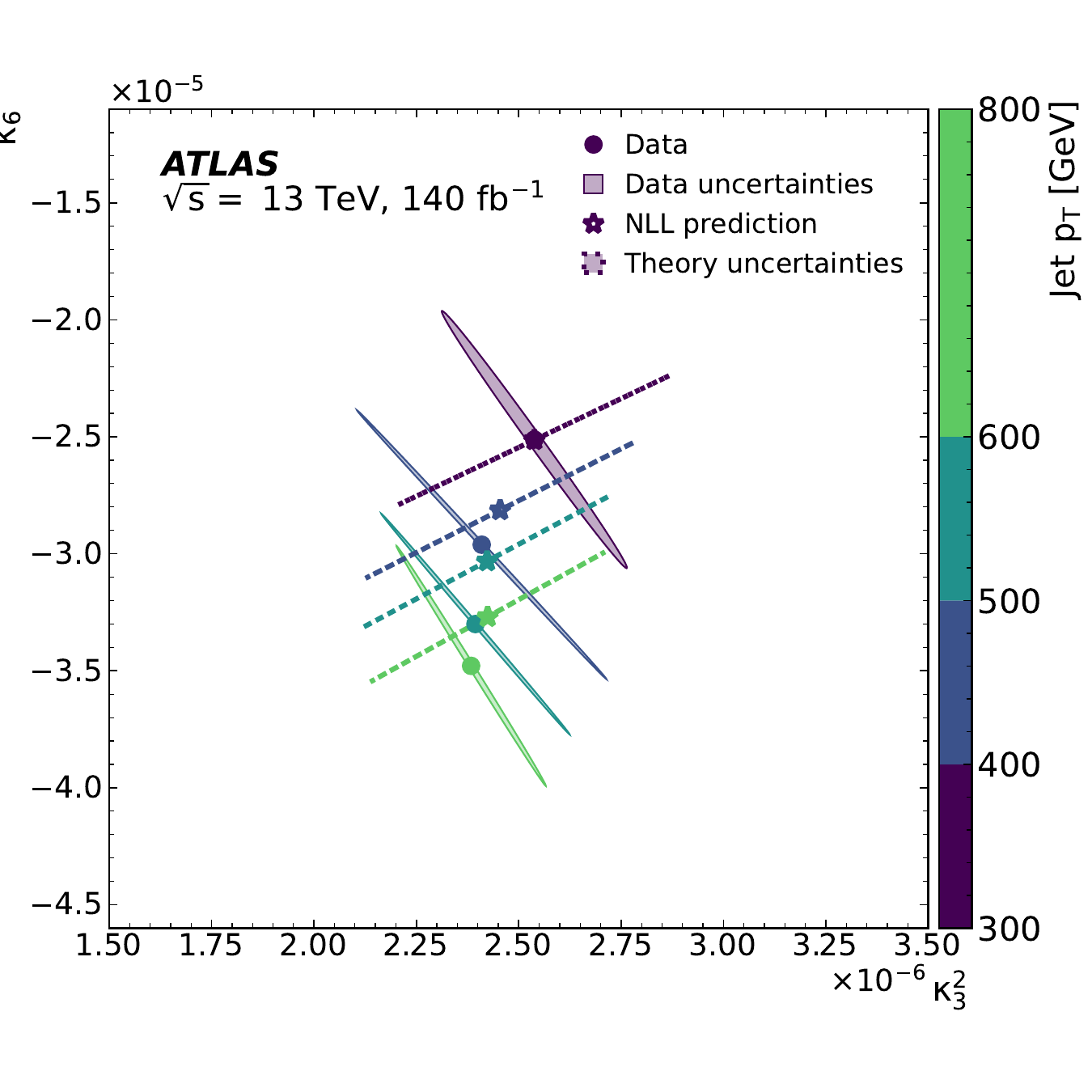}\label{fig:flow:d}}
\caption{Non-trivial RG flow relationship between higher-order cumulants and products of lower-order ones coming from the non-linearity of the track function evolution. The relationship between (a) $\kappa_4$ and $\kappa_2 ^2$, (b) $\kappa_5$ and $\kappa_2\kappa_3$, (c) $\kappa_6$ and $\kappa_2 ^3$, and (d) $\kappa_6$ and $\kappa_3 ^2$ are shown for different bins of the particle-level jet \pt.
The ellipses correspond to the 68\% confidence regions based on the combined statistical and systematic uncertainty (data with solid boundaries) and scale variations (theory with dotted boundaries).
}
\label{fig:flow}
\end{figure}

The extracted moments of the \rcharged distribution can be expressed in terms of the cumulants of the distribution, $\kappa_n$, which can be written in terms of the central moments $\mu_n =\text{E}[(X-\text{E}[X])^n]$, as $\kappa_2=\mu_2$, $\kappa_3=\mu_3$, $\kappa_4=\mu_4-3\mu_2^2$, $\kappa_5=\mu_5-10 \mu_3 \mu_2$, and $\kappa_6=\mu_6-15\mu_4\mu_2- 10\mu_3^2 +30\mu_2^3$~\cite{Li_2022,Lee:2023xzv}.
The energy dependence of the relationship between pairs of these cumulants is theoretically determined by non-trivial renormalization group flows (RG flows) that govern the scale dependence of correlations in the hadronization process.
Physically, the track function depicts not only the energy fraction of the initial hard parton converted into all the charged hadrons, but also the correlations among them~\cite{Chang:2013rca}.
One prediction of this theory is that as the energy increases, the cumulants should converge towards some fixed values that depend on the quark-/gluon-initiated jet composition (the cumulants for demixed quark-/gluon-initiated jet \rcharged distributions should converge to 0).

Figure~\ref{fig:flow} shows the relationship between selected pairs of cumulants extracted from the unfolded data distributions and compared to an analytical QCD prediction at next-to-leading-logarithm (NLL) of the RG flow provided by the authors of Ref.~\cite{Lee:2023xzv}.
The NLL predictions perturbatively evolve the track function distribution using data in the lowest \pt bin as the initial condition, meaning that the direction of flow can be compared, but not each measurement individually.
Due to the strong correlations in uncertainties between the cumulants, the 68\% confidence regions for each measurement have a highly elliptical shape for both theory and experiment.
Binning corrections for the cumulants are derived using the same method as for the moments.
While these corrections tend to be close to 1, they can be much larger or smaller for some of the higher cumulants.

Comparing the shapes of the flows between data and theory provides a direct test of non-linear renormalization group flows during jet formation.
The results in Figures~\ref{fig:flow:a}--\ref{fig:flow:d} show good data-theory agreement and are consistent with the picture of flow towards a fixed point.
This is seen by the fact that the moments evolve more slowly at higher momentum scales.
The convergence may be better tested at even higher momentum scales.
\FloatBarrier

\section{Conclusions}

This paper presents a measurement of the differential cross-section of the transverse momentum fraction \rcharged carried by charged hadrons of jets, and the scale dependence of its moments.
The measurement is performed on an inclusive selections of dijet events, using a dataset corresponding to an integrated luminosity of $140$~fb$^{-1}$ of proton--proton collisions at $\sqrt{s}=13$~\TeV recorded with the ATLAS detector at the LHC.
Jets are reconstructed using the anti-$k_t$ algorithm with $R=0.4$, and their associated charged-particle tracks are used to measure the unfolded single-differential cross-section of \rcharged, the observable of interest.
The moments of this distribution are extracted, which are used to construct the lower-order cumulants and their relationships.
This measurement of the \rcharged distribution enables the determination of universal track functions, which are a necessary component in providing theoretical calculations of jet substructure observables using charged particles.
This paves the way towards theoretical precision for a new class of observables that will enable further studies of jet formation and the strong interaction.

\section*{Acknowledgements}

%
%

%
%

We thank CERN for the very successful operation of the LHC and its injectors, as well as the support staff at
CERN and at our institutions worldwide without whom ATLAS could not be operated efficiently.

The crucial computing support from all WLCG partners is acknowledged gratefully, in particular from CERN, the ATLAS Tier-1 facilities at TRIUMF/SFU (Canada), NDGF (Denmark, Norway, Sweden), CC-IN2P3 (France), KIT/GridKA (Germany), INFN-CNAF (Italy), NL-T1 (Netherlands), PIC (Spain), RAL (UK) and BNL (USA), the Tier-2 facilities worldwide and large non-WLCG resource providers. Major contributors of computing resources are listed in Ref.~\cite{ATL-SOFT-PUB-2025-001}.

We gratefully acknowledge the support of ANPCyT, Argentina; YerPhI, Armenia; ARC, Australia; BMWFW and FWF, Austria; ANAS, Azerbaijan; CNPq and FAPESP, Brazil; NSERC, NRC and CFI, Canada; CERN; ANID, Chile; CAS, MOST and NSFC, China; Minciencias, Colombia; MEYS CR, Czech Republic; DNRF and DNSRC, Denmark; IN2P3-CNRS and CEA-DRF/IRFU, France; SRNSFG, Georgia; BMBF, HGF and MPG, Germany; GSRI, Greece; RGC and Hong Kong SAR, China; ISF and Benoziyo Center, Israel; INFN, Italy; MEXT and JSPS, Japan; CNRST, Morocco; NWO, Netherlands; RCN, Norway; MNiSW, Poland; FCT, Portugal; MNE/IFA, Romania; MSTDI, Serbia; MSSR, Slovakia; ARIS and MVZI, Slovenia; DSI/NRF, South Africa; MICIU/AEI, Spain; SRC and Wallenberg Foundation, Sweden; SERI, SNSF and Cantons of Bern and Geneva, Switzerland; NSTC, Taipei; TENMAK, T\"urkiye; STFC/UKRI, United Kingdom; DOE and NSF, United States of America.

Individual groups and members have received support from BCKDF, CANARIE, CRC and DRAC, Canada; CERN-CZ, FORTE and PRIMUS, Czech Republic; COST, ERC, ERDF, Horizon 2020, ICSC-NextGenerationEU and Marie Sk{\l}odowska-Curie Actions, European Union; Investissements d'Avenir Labex, Investissements d'Avenir Idex and ANR, France; DFG and AvH Foundation, Germany; Herakleitos, Thales and Aristeia programmes co-financed by EU-ESF and the Greek NSRF, Greece; BSF-NSF and MINERVA, Israel; NCN and NAWA, Poland; La Caixa Banking Foundation, CERCA Programme Generalitat de Catalunya and PROMETEO and GenT Programmes Generalitat Valenciana, Spain; G\"{o}ran Gustafssons Stiftelse, Sweden; The Royal Society and Leverhulme Trust, United Kingdom.

In addition, individual members wish to acknowledge support from Armenia: Yerevan Physics Institute (FAPERJ); CERN: European Organization for Nuclear Research (CERN PJAS); Chile: Agencia Nacional de Investigaci\'on y Desarrollo (FONDECYT 1230812, FONDECYT 1230987, FONDECYT 1240864); China: Chinese Ministry of Science and Technology (MOST-2023YFA1605700, MOST-2023YFA1609300), National Natural Science Foundation of China (NSFC - 12175119, NSFC 12275265, NSFC-12075060); Czech Republic: Czech Science Foundation (GACR - 24-11373S), Ministry of Education Youth and Sports (FORTE CZ.02.01.01/00/22\_008/0004632), PRIMUS Research Programme (PRIMUS/21/SCI/017); EU: H2020 European Research Council (ERC - 101002463); European Union: European Research Council (ERC - 948254, ERC 101089007), Horizon 2020 Framework Programme (MUCCA - CHIST-ERA-19-XAI-00), European Union, Future Artificial Intelligence Research (FAIR-NextGenerationEU PE00000013), Italian Center for High Performance Computing, Big Data and Quantum Computing (ICSC, NextGenerationEU); France: Agence Nationale de la Recherche (ANR-20-CE31-0013, ANR-21-CE31-0013, ANR-21-CE31-0022, ANR-22-EDIR-0002), Investissements d'Avenir Labex (ANR-11-LABX-0012); Germany: Baden-Württemberg Stiftung (BW Stiftung-Postdoc Eliteprogramme), Deutsche Forschungsgemeinschaft (DFG - 469666862, DFG - CR 312/5-2); Italy: Istituto Nazionale di Fisica Nucleare (ICSC, NextGenerationEU), Ministero dell'Università e della Ricerca (PRIN - 20223N7F8K - PNRR M4.C2.1.1); Japan: Japan Society for the Promotion of Science (JSPS KAKENHI JP22H01227, JSPS KAKENHI JP22H04944, JSPS KAKENHI JP22KK0227, JSPS KAKENHI JP23KK0245); Netherlands: Netherlands Organisation for Scientific Research (NWO Veni 2020 - VI.Veni.202.179); Norway: Research Council of Norway (RCN-314472); Poland: Ministry of Science and Higher Education (IDUB AGH, POB8, D4 no 9722), Polish National Agency for Academic Exchange (PPN/PPO/2020/1/00002/U/00001), Polish National Science Centre (NCN 2021/42/E/ST2/00350, NCN OPUS nr 2022/47/B/ST2/03059, NCN UMO-2019/34/E/ST2/00393, NCN \& H2020 MSCA 945339, UMO-2020/37/B/ST2/01043, UMO-2021/40/C/ST2/00187, UMO-2022/47/O/ST2/00148, UMO-2023/49/B/ST2/04085, UMO-2023/51/B/ST2/00920); Slovenia: Slovenian Research Agency (ARIS grant J1-3010); Spain: Generalitat Valenciana (Artemisa, FEDER, IDIFEDER/2018/048), Ministry of Science and Innovation (MCIN \& NextGenEU PCI2022-135018-2, MICIN \& FEDER PID2021-125273NB, RYC2019-028510-I, RYC2020-030254-I, RYC2021-031273-I, RYC2022-038164-I), PROMETEO and GenT Programmes Generalitat Valenciana (CIDEGENT/2019/027); Sweden: Carl Trygger Foundation (Carl Trygger Foundation CTS 22:2312), Swedish Research Council (Swedish Research Council 2023-04654, VR 2018-00482, VR 2022-03845, VR 2022-04683, VR 2023-03403, VR grant 2021-03651), Knut and Alice Wallenberg Foundation (KAW 2018.0157, KAW 2018.0458, KAW 2019.0447, KAW 2022.0358); Switzerland: Swiss National Science Foundation (SNSF - PCEFP2\_194658); United Kingdom: Leverhulme Trust (Leverhulme Trust RPG-2020-004), Royal Society (NIF-R1-231091); United States of America: U.S. Department of Energy (ECA DE-AC02-76SF00515), Neubauer Family Foundation.

%
%


%

\clearpage
\appendix
\part*{Appendix}
\addcontentsline{toc}{part}{Appendix}

\section{Quark and gluon interpretation}
\label{app:qgInterpretation}

A statistical demixing of the quark- and gluon-like cross-sections is performed, following the demixing strategy outlined in Ref.~\cite{STDM-2017-16}.
Since the shape of the \rcharged distribution for a given jet \pt depends on the flavor of the initiating parton but not $\eta$, the quark and gluon distributions may be inferred from the measurements of the central and forward distributions using: $h_{i}^{F} = f_{Q}^{F} h_{i}^{Q} + f_{G}^{F} h_{i}^{G}$ and $h_{i}^{C} = f_{Q}^{C} h_{i}^{Q} + f_{G}^{C} h_{i}^{G}$,
where $h_i$ is one bin of the histogram, $F$ and $C$ denote the forward and central regions, and $Q$ and $G$ represent quark or gluon respectively.

The quark and gluon fractions for the more forward and more central jets are obtained from the nominal \PYTHIA8 MC event sample,\footnote{The jet flavor is determined by the highest energy parton associated to the jet.} and are shown in Figure~\ref{fig:objects:frac}.
By taking these fractions from an MC prediction, this results in a model-dependence, and so an additional uncertainty is applied to account for differences in the predicted quark and gluon fractions across different theoretical models.

\begin{figure}[htbp]
\centering
\subfigure{\includegraphics[width=0.6\textwidth]{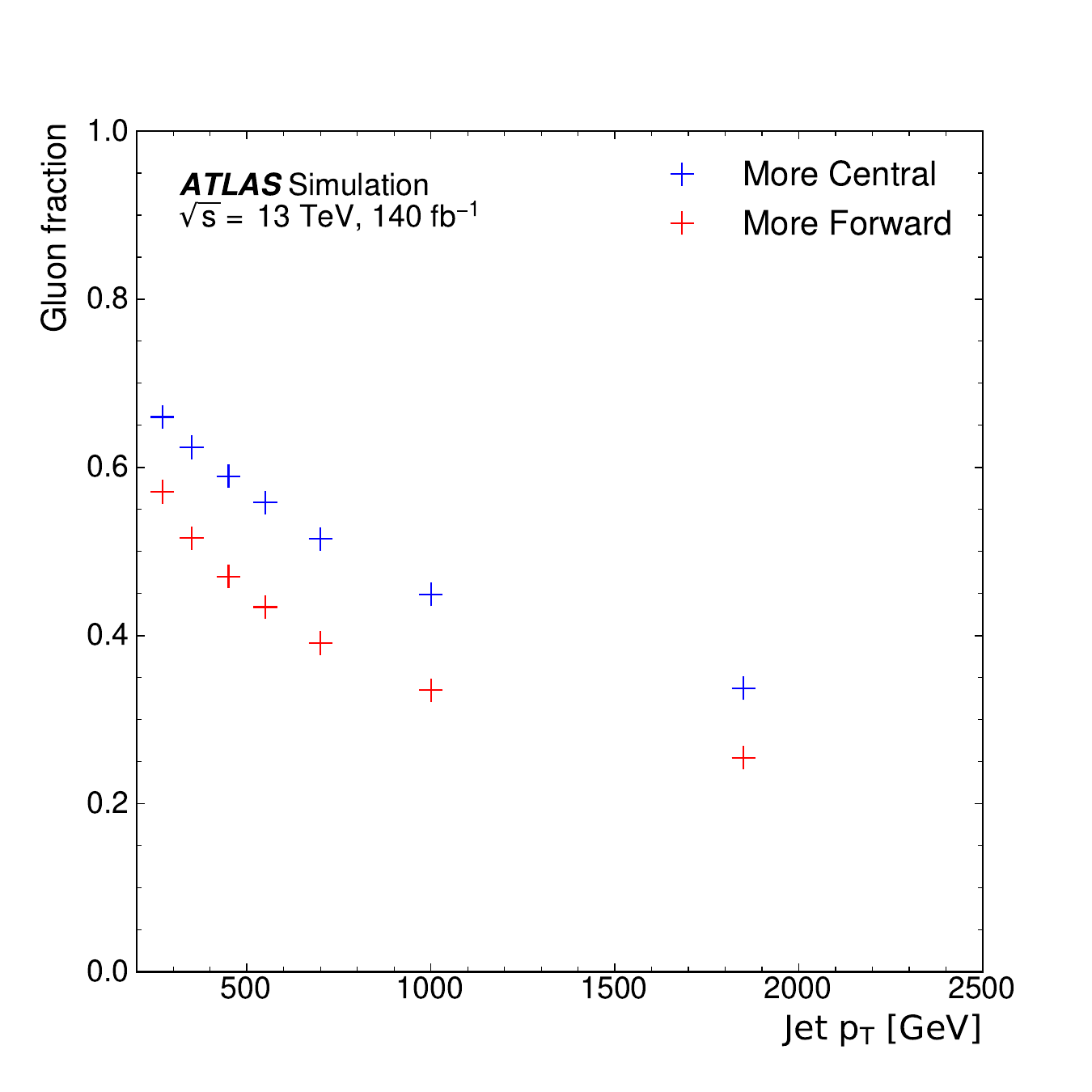}}
\caption{The gluon fraction obtained from the nominal \PYTHIA8 MC event sample for the more forward and more central of the two dijets as a function of the jet \pt. The cross markers do not include uncertainties in this plot.}
\label{fig:objects:frac}
\end{figure}

The demixed quark and gluon interpretation of $r_q$ are shown in Figure~\ref{fig:xsecqg} for two representative jet \pt bins.
In general, the agreement between data and MC is similar to the agreement with the measured cross-sections, with larger discrepancies for the gluon extraction than for the quark extraction.
The statistically demixed gluon-initiated jets have a slightly narrower \rcharged distribution than that found for quark-initiated jets due to the larger color factor, but the overall distributions are very similar.
The central moments of the \rcharged distribution are shown in Figure~\ref{fig:momentqg} for the extracted quark and gluon cross-sections.
All six moments extracted from the demixed \rcharged distributions are systematically higher for quark-initiated jets than for gluon-initiated jets.

\begin{figure}[t!]
\centering
\subfigure[]{\includegraphics[width=0.49\textwidth]{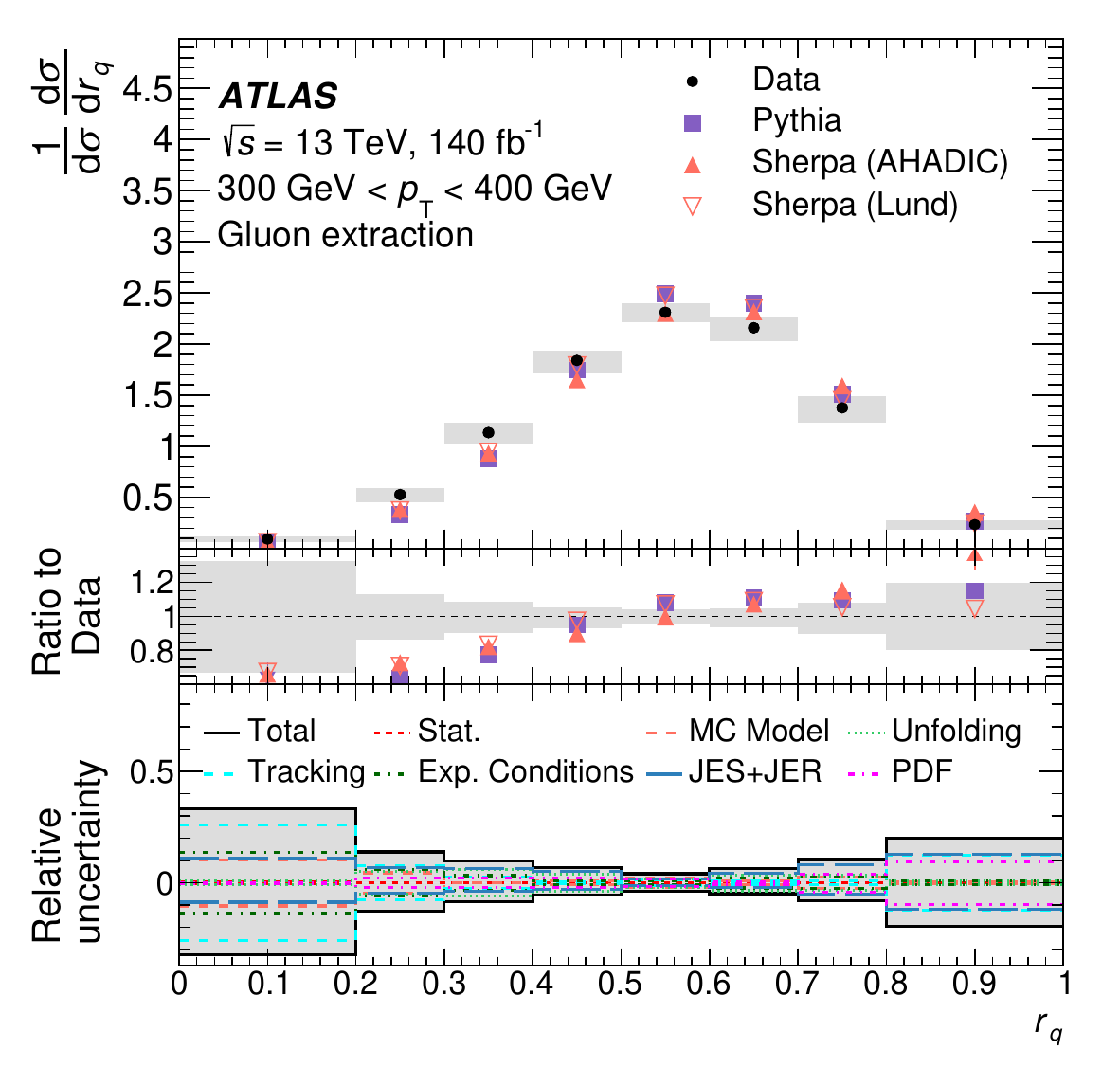}}
\subfigure[]{\includegraphics[width=0.49\textwidth]{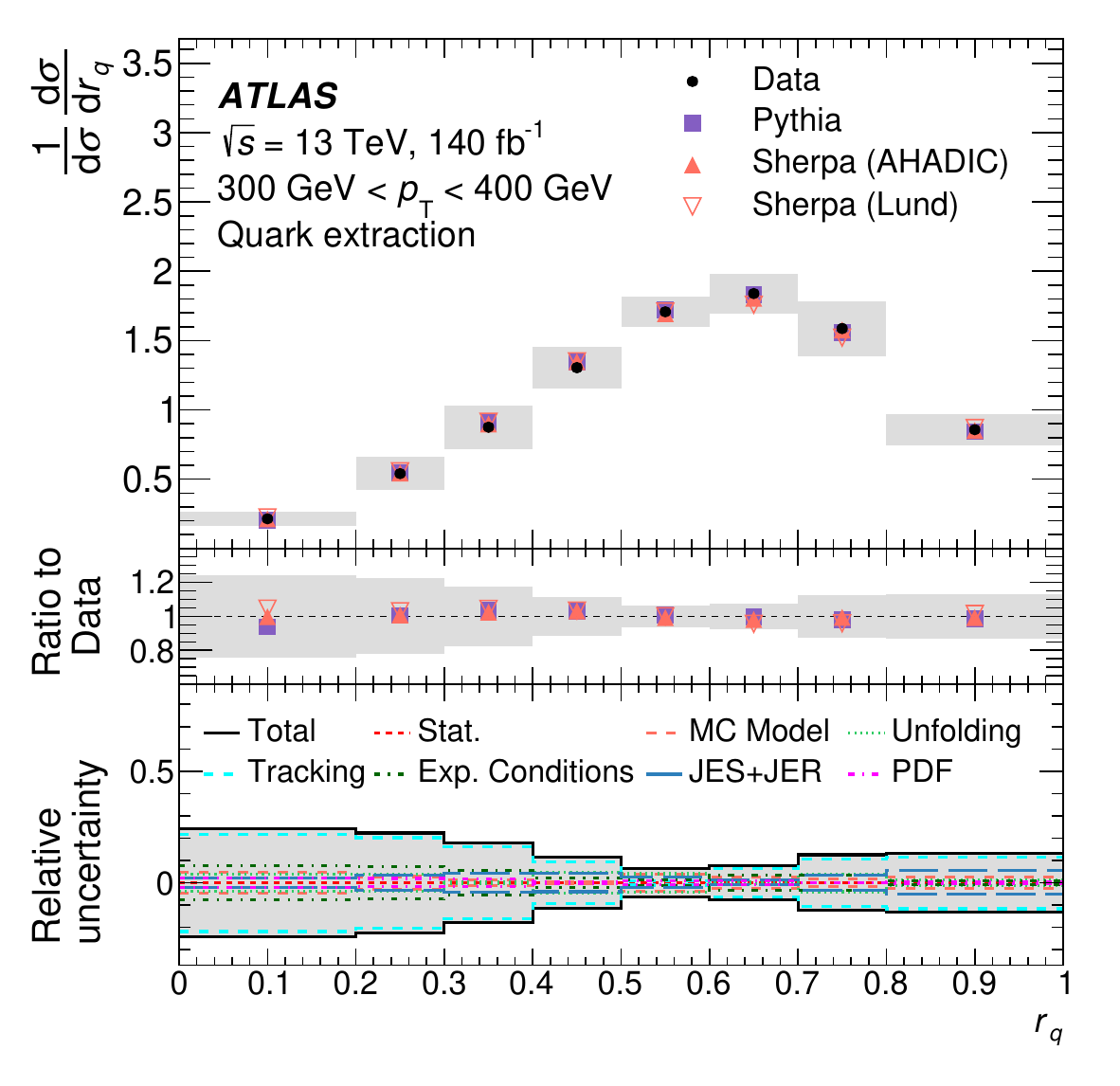}}
\subfigure[]{\includegraphics[width=0.49\textwidth]{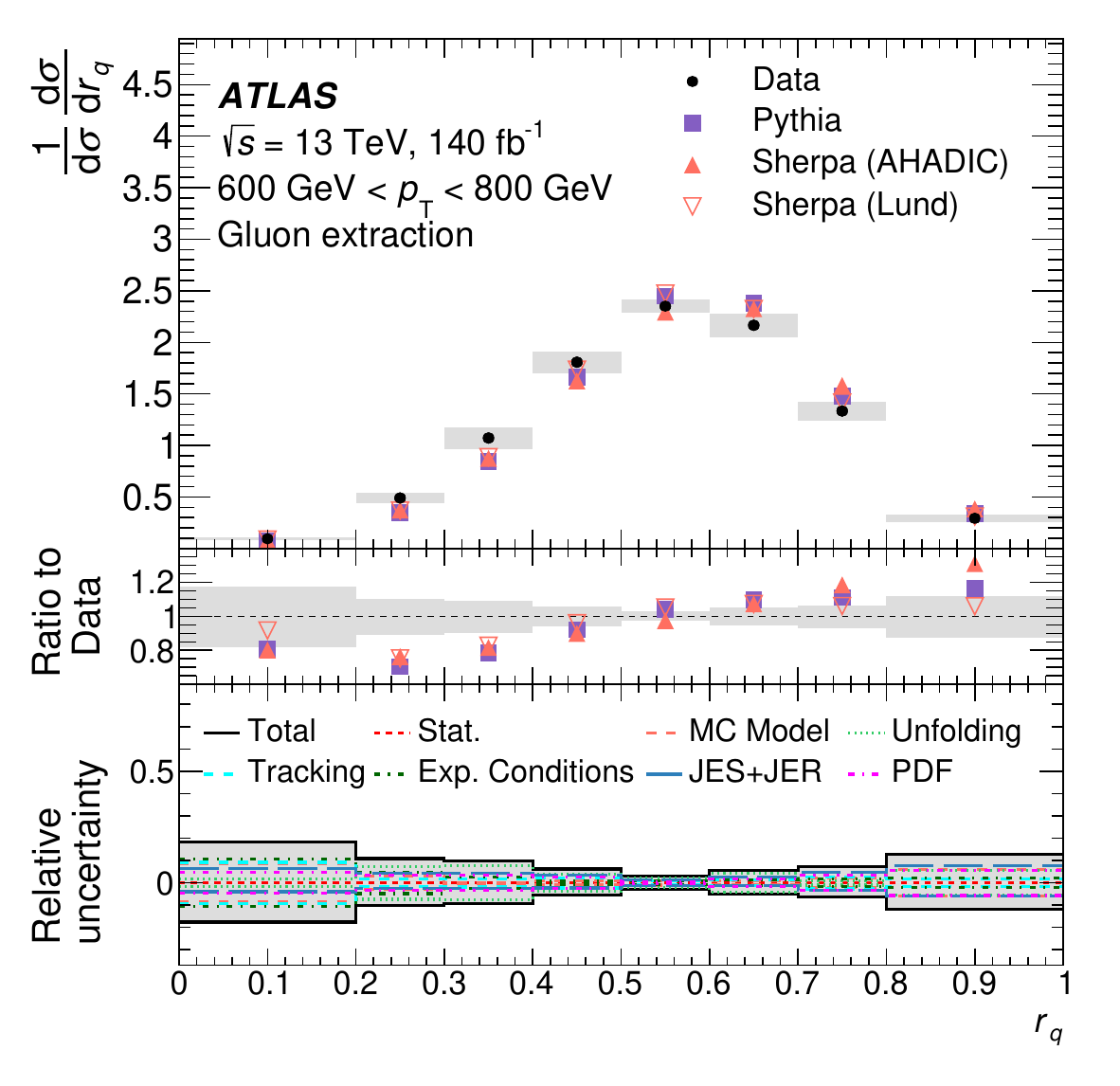}}
\subfigure[]{\includegraphics[width=0.49\textwidth]{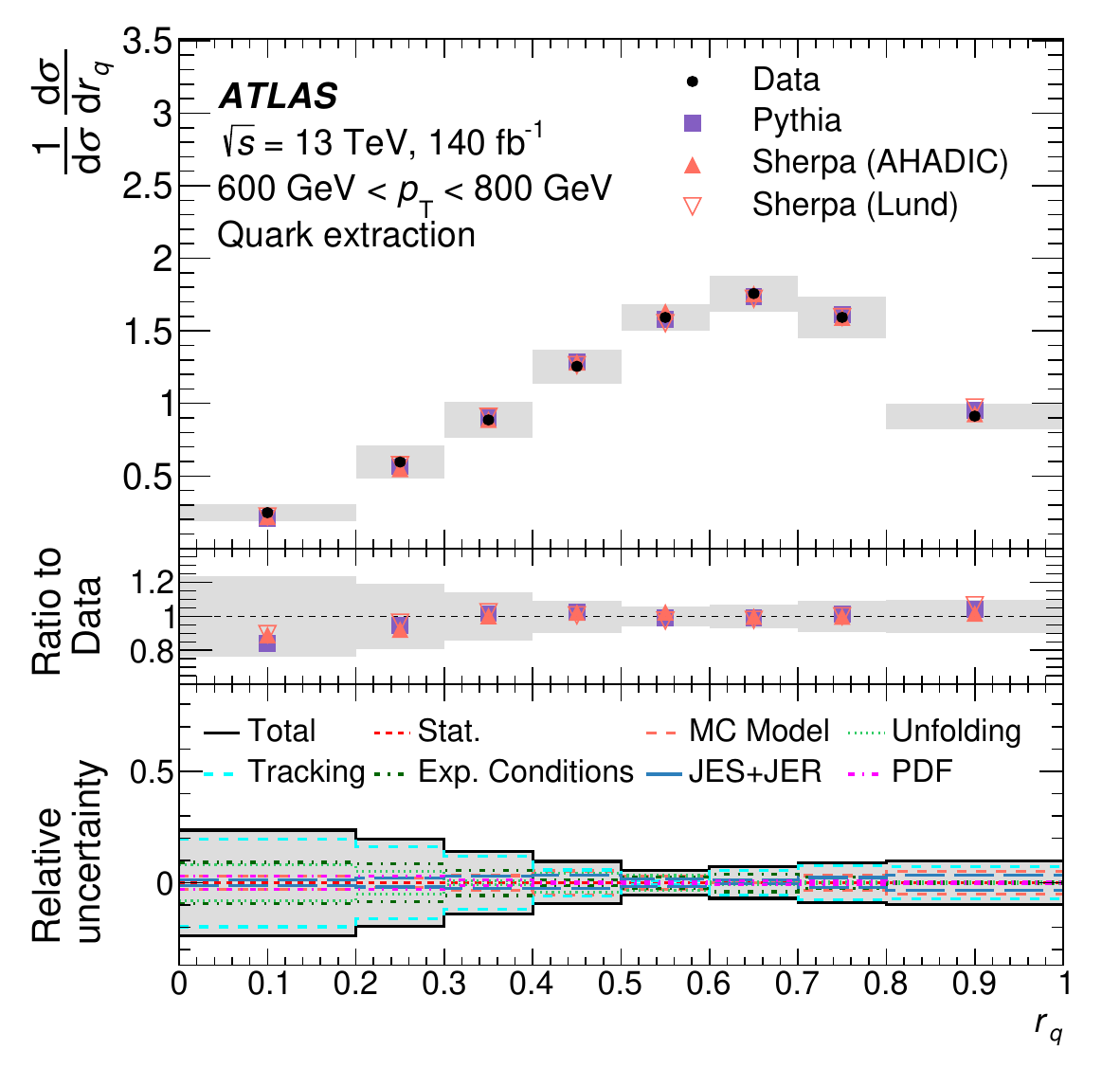}}
\caption{The demixed (a, c) gluon and (b, d) quark interpretation of the unfolded \rcharged cross-sections for (a, b) $300~\GeV < \pt < 400~\GeV$, and (c, d) $600~\GeV < \pt < 800~\GeV$ for data compared to predictions from several MC generators. The grey uncertainty band shows the combined statistical and systematic uncertainties on the measurement.
The bottom panel shows the breakdown of the uncertainties grouped by their sources.
In particular, the `Exp. Conditions' group includes uncertainties related to the overlaid pileup events and potential biases due to differences between disabled modules of the Tile calorimeter.
}
\label{fig:xsecqg}
\end{figure}

\begin{figure}[t!] %
\centering
\subfigure[]{\includegraphics[width=0.49\textwidth]{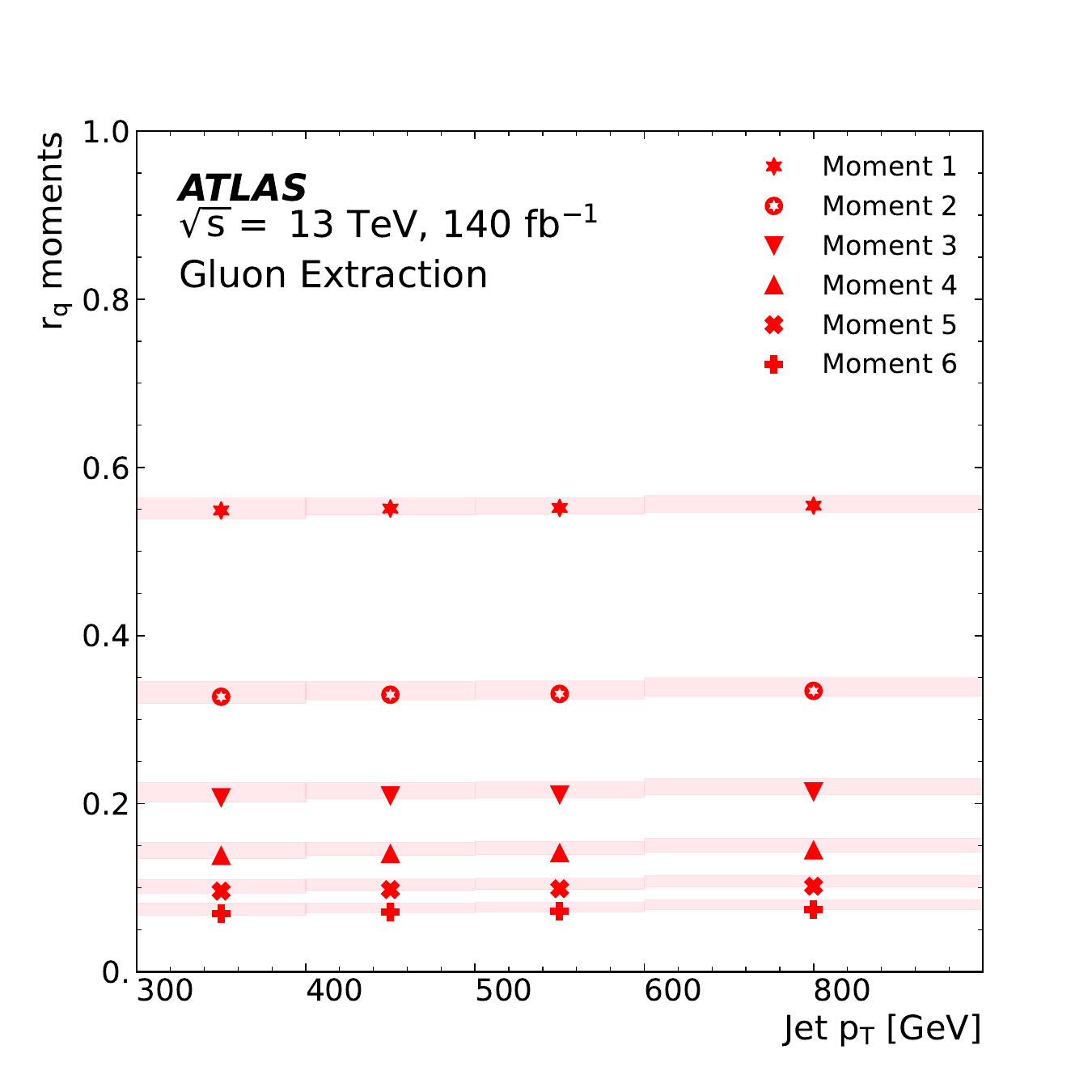}}
\subfigure[]{\includegraphics[width=0.49\textwidth]{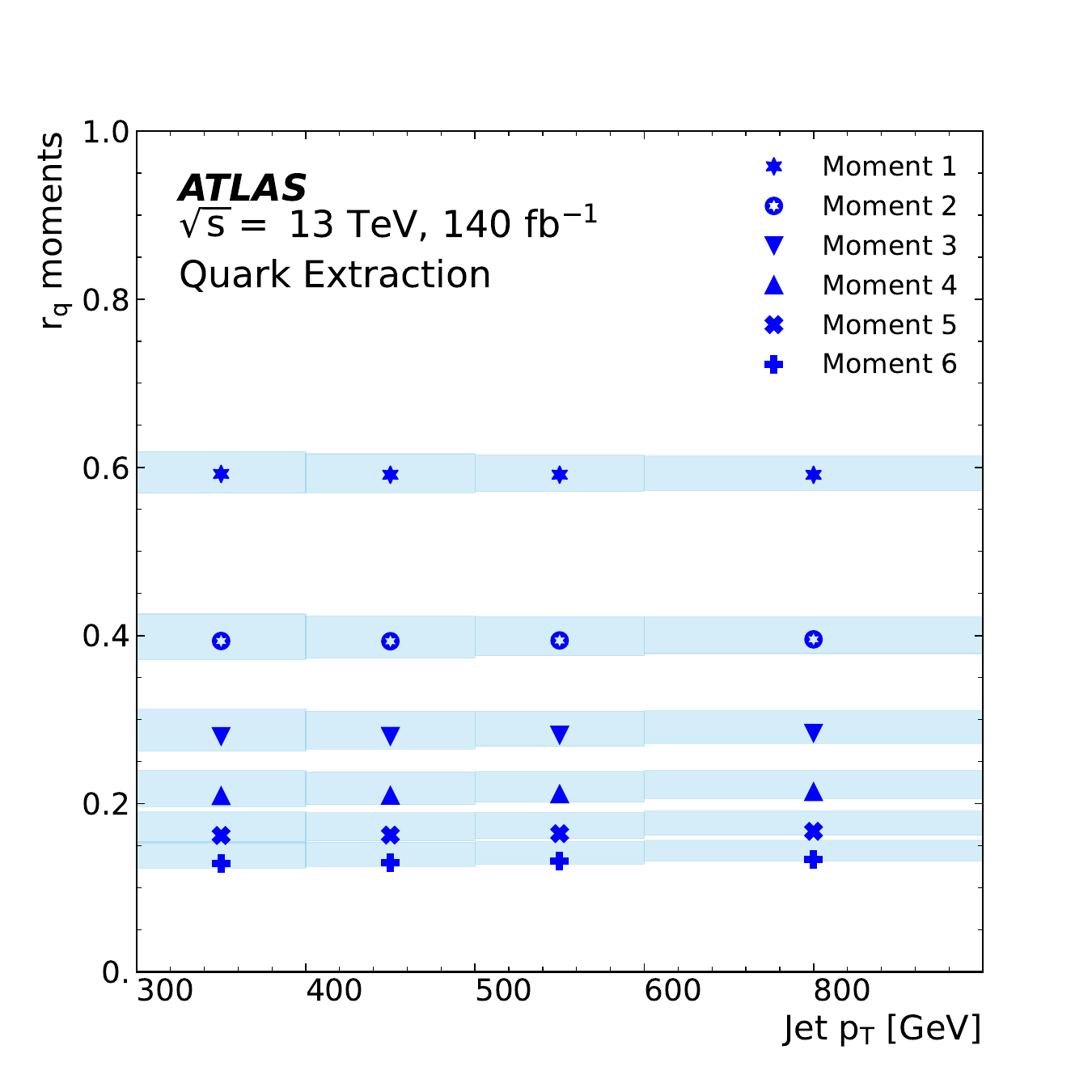}}
\caption{The first six moments of the demixed (a) gluon and (b) quark \rcharged distributions in jet \pt bins. The uncertainty bands show the combined statistical and systematic uncertainties on the measurement.
}
\label{fig:momentqg}
\end{figure}

\printbibliography

@article{ATLAS:2024vdo,
	archiveprefix = {arXiv},
	author = "{ATLAS Collaboration}",
	collaboration = {ATLAS},
	eprint = {2404.06335},
	month = {4},
	primaryclass = {hep-ex},
	reportnumber = {CERN-EP-2024-100},
	title = {{Software and computing for Run 3 of the ATLAS experiment at the LHC}},
	year = {2024}}

@article{STDM-2020-31,
	archiveprefix = {arXiv},
	author = {{ATLAS Collaboration}},
	eprint = {2407.10879},
	month = {7},
	primaryclass = {hep-ex},
	reportnumber = {CERN-EP-2024-169},
	title = {{Measurement of the Lund jet plane in hadronic decays of top quarks and W bosons with the ATLAS detector}},
	year = {2024}}

@article{ALICE:2019ykw,
	archiveprefix = {arXiv},
	author = {{ALICE Collaboration}},
	doi = {10.1016/j.physletb.2020.135227},
	eprint = {1905.02512},
	journal = {Phys. Lett. B},
	pages = {135227},
	primaryclass = {nucl-ex},
	reportnumber = {CERN-EP-2019-087},
	title = {{Exploration of jet substructure using iterative declustering in pp and Pb\textendash{}Pb collisions at LHC energies}},
	volume = {802},
	year = {2020}}

@article{ALICE:2021njq,
	archiveprefix = {arXiv},
	author = {{ALICE Collaboration}},
	doi = {10.1007/JHEP05(2022)061},
	eprint = {2107.11303},
	journal = {JHEP},
	pages = {061},
	primaryclass = {nucl-ex},
	reportnumber = {CERN-EP-2021-145},
	title = {{Measurements of the groomed and ungroomed jet angularities in pp collisions at $ \sqrt{s} $ = 5.02 TeV}},
	volume = {05},
	year = {2022}}

@article{ALICE:2021aqk,
	author = "{ALICE Collaboration}",
	title = "{Direct observation of the dead-cone effect in quantum chromodynamics}",
	eprint = "2106.05713",
	archivePrefix = "arXiv",
	primaryClass = "nucl-ex",
	reportNumber = "CERN-EP-2021-107",
	doi = "10.1038/s41586-022-04572-w",
	journal = "Nature",
	volume = "605",
	number = "7910",
	pages = "440--446",
	year = "2022",
    related = "ALICE:2021aqk-err",
    relatedstring = "Erratum:",
}

@article{LHCb:2017llq,
	author = "{LHCb Collaboration}",
	title = "{Study of J/\ensuremath{\psi} Production in Jets}",
	eprint = "1701.05116",
	archivePrefix = "arXiv",
	primaryClass = "hep-ex",
	reportNumber = "LHCB-PAPER-2016-064, CERN-EP-2017-006",
	doi = "10.1103/PhysRevLett.118.192001",
	journal = "Phys. Rev. Lett.",
	volume = "118",
	number = "19",
	pages = "192001",
	year = "2017"
}

@article{LHCb:2019qoc,
	author = "{LHCb Collaboration}",
	title = "{Measurement of Charged Hadron Production in $Z$-Tagged Jets in Proton-Proton Collisions at $ \sqrt{s} = 8$ TeV}",
	eprint = "1904.08878",
	archivePrefix = "arXiv",
	primaryClass = "hep-ex",
	reportNumber = "CERN-EP-2019-065, LHCb-PAPER-2019-012",
	doi = "10.1103/PhysRevLett.123.232001",
	journal = "Phys. Rev. Lett.",
	volume = "123",
	number = "23",
	pages = "232001",
	year = "2019"
}

@article{Chen:2020vvp,
	archiveprefix = {arXiv},
	author = {Chen, Hao and Moult, Ian and Zhang, XiaoYuan and Zhu, Hua Xing},
	doi = {10.1103/PhysRevD.102.054012},
	eprint = {2004.11381},
	journal = {Phys. Rev. D},
	number = {5},
	pages = {054012},
	primaryclass = {hep-ph},
	title = {{Rethinking jets with energy correlators: Tracks, resummation, and analytic continuation}},
	volume = {102},
	year = {2020}}

@article{Altarelli:1977zs,
	author = {Altarelli, Guido and Parisi, G.},
	doi = {10.1016/0550-3213(77)90384-4},
	journal = {Nucl. Phys. B},
	pages = {298--318},
	reportnumber = {LPTENS-77-6},
	title = {{Asymptotic Freedom in Parton Language}},
	volume = {126},
	year = {1977}}

@article{Dokshitzer:1977sg,
	author = {Dokshitzer, Yuri L.},
	journal = {Sov. Phys. JETP},
	pages = {641--653},
	title = {{Calculation of the Structure Functions for Deep Inelastic Scattering and e+ e- Annihilation by Perturbation Theory in Quantum Chromodynamics.}},
	volume = {46},
	year = {1977}}

@article{Gribov:1972ri,
	author = {Gribov, V. N. and Lipatov, L. N.},
	journal = {Sov. J. Nucl. Phys.},
	pages = {438--450},
	reportnumber = {IPTI-381-71},
	title = {{Deep inelastic e p scattering in perturbation theory}},
	volume = {15},
	year = {1972}}

@article{Chen:2022muj,
	archiveprefix = {arXiv},
	author = {Chen, Hao and Jaarsma, Max and Li, Yibei and Moult, Ian and Waalewijn, Wouter J. and Zhu, Hua Xing},
	eprint = {2210.10061},
	month = {10},
	primaryclass = {hep-ph},
	title = {{Collinear Parton Dynamics Beyond DGLAP}},
	year = {2022}}

@inproceedings{Andreassen:2021zzk,
	archiveprefix = {arXiv},
	author = {Andreassen, Anders and Komiske, Patrick T. and Metodiev, Eric M. and Nachman, Benjamin and Suresh, Adi and Thaler, Jesse},
	booktitle = {{9th International Conference on Learning Representations}},
	eprint = {2105.04448},
	month = {5},
	primaryclass = {stat.ML},
	title = {{Scaffolding Simulations with Deep Learning for High-dimensional Deconvolution}},
	year = {2021}}

@article{Li_2022,
	author = "Li, Yibei and Moult, Ian and van Velzen, Solange Schrijnder and Waalewijn, Wouter J. and Zhu, Hua Xing",
eprint = "2108.01674",
archivePrefix = "arXiv",
primaryClass = "hep-ph",
	doi = {10.1103/physrevlett.128.182001},
	journal = {Phys. Rev. Lett.},
	month = {5},
	number = {18},
	publisher = {American Physical Society ({APS})},
	title = {Extending Precision Perturbative {QCD} with Track Functions},
	volume = {128},
	year = 2022,
pages = "182001",
}

@article{Lucy:1974yx,
	author = {Lucy, L. B.},
	doi = {10.1086/111605},
	journal = {Astron. J.},
	pages = {745--754},
	title = {{An iterative technique for the rectification of observed distributions}},
	volume = {79},
	year = {1974}}

@article{ALargeIonColliderExperiment:2021mqf,
	archiveprefix = {arXiv},
	author = {{ALICE Collaboration}},
	doi = {10.1103/PhysRevLett.128.102001},
	eprint = {2107.12984},
	journal = {Phys. Rev. Lett.},
	number = {10},
	pages = {102001},
	primaryclass = {nucl-ex},
	reportnumber = {CERN-EP-2021-151},
	title = {{Measurement of the Groomed Jet Radius and Momentum Splitting Fraction in pp and Pb$-$Pb Collisions at $\sqrt{s_{NN}} = 5.02$~TeV}},
	volume = {128},
	year = {2022}}

@article{Richardson:72,
	author = {William Hadley Richardson},
	doi = {10.1364/JOSA.62.000055},
	journal = {J. Opt. Soc. Am.},
	number = {1},
	pages = {55--59},
	publisher = {OSA},
	title = {Bayesian-Based Iterative Method of Image Restoration},
	volume = {62},
	year = {1972}}

@article{Chang:2013rca,
	archiveprefix = {arXiv},
	author = {Chang, Hsi-Ming and Procura, Massimiliano and Thaler, Jesse and Waalewijn, Wouter J.},
	doi = {10.1103/PhysRevLett.111.102002},
	eprint = {1303.6637},
	journal = {Phys. Rev. Lett.},
	pages = {102002},
	primaryclass = {hep-ph},
	reportnumber = {MIT-CTP-4449, MIT--CTP-4449},
	title = {{Calculating Track-Based Observables for the LHC}},
	volume = {111},
	year = {2013}}

@article{Chang:2013iba,
	archiveprefix = {arXiv},
	author = {Chang, Hsi-Ming and Procura, Massimiliano and Thaler, Jesse and Waalewijn, Wouter J.},
	doi = {10.1103/PhysRevD.88.034030},
	eprint = {1306.6630},
	journal = {Phys. Rev. D},
	pages = {034030},
	primaryclass = {hep-ph},
	reportnumber = {MIT--CTP-4476},
	title = {{Calculating track thrust with track functions}},
	volume = {88},
	year = {2013}}

@article{1911.09107,
	archiveprefix = {arXiv},
	author = {{A. Andreassen, E. Metodiev, P. Komiske, B. Nachman, and J. Thaler}},
	doi = {10.1103/PhysRevLett.124.182001},
	eprint = {1911.09107},
	journal = {Phys. Rev. Lett.},
	pages = {182001},
	primaryclass = {hep-ph},
	title = {{OmniFold: A Method to Simultaneously Unfold All Observables}},
	volume = {124},
	year = {2020}}

@article{Jaarsma:2023ell,
	archiveprefix = {arXiv},
	author = {Jaarsma, Max and Li, Yibei and Moult, Ian and Waalewijn, Wouter J. and Zhu, H. X.},
	eprint = {2307.15739},
	month = {7},
	primaryclass = {hep-ph},
	title = {{Energy Correlators on Tracks: Resummation and Non-Perturbative Effects}},
	year = {2023}}

@article{Chen:2022pdu,
	archiveprefix = {arXiv},
	author = {Chen, Hao and Jaarsma, Max and Li, Yibei and Moult, Ian and Waalewijn, Wouter J. and Zhu, H. X.},
	doi = {10.1007/JHEP07(2023)185},
	eprint = {2210.10058},
	journal = {JHEP},
	pages = {185},
	primaryclass = {hep-ph},
	title = {{Multi-collinear splitting kernels for track function evolution}},
	volume = {07},
	year = {2023}}

@article{Jaarsma:2022kdd,
	archiveprefix = {arXiv},
	author = {Jaarsma, Max and Li, Yibei and Moult, Ian and Waalewijn, Wouter and Zhu, Hua X.},
	doi = {10.1007/JHEP06(2022)139},
	eprint = {2201.05166},
	journal = {JHEP},
	pages = {139},
	primaryclass = {hep-ph},
	title = {{Renormalization group flows for track function moments}},
	volume = {06},
	year = {2022}}

@article{Li:2021zcf,
	archiveprefix = {arXiv},
	author = {Li, Yibei and Moult, Ian and van Velzen, Solange Schrijnder and Waalewijn, Wouter J. and Zhu, H. X.},
	doi = {10.1103/PhysRevLett.128.182001},
	eprint = {2108.01674},
	journal = {Phys. Rev. Lett.},
	number = {18},
	pages = {182001},
	primaryclass = {hep-ph},
	title = {{Extending Precision Perturbative QCD with Track Functions}},
	volume = {128},
	year = {2022}}

@article{Larkoski:2017jix,
	archiveprefix = {arXiv},
	author = {Larkoski, Andrew J. and Moult, Ian and Nachman, Benjamin},
	doi = {10.1016/j.physrep.2019.11.001},
	eprint = {1709.04464},
	journal = {Phys. Rept.},
	pages = {1--63},
	primaryclass = {hep-ph},
	title = {{Jet substructure at the Large Hadron Collider: A review of recent advances in theory and machine learning}},
	volume = {841},
	year = {2020}}

@article{Kogler:2018hem,
	archiveprefix = {arXiv},
	author = {Kogler, Roman and others},
	doi = {10.1103/RevModPhys.91.045003},
	eprint = {1803.06991},
	journal = {Rev. Mod. Phys.},
	number = {4},
	pages = {045003},
	primaryclass = {hep-ex},
	reportnumber = {FERMILAB-PUB-18-123-PPD},
	title = {{Jet substructure at the Large Hadron Collider}},
	volume = {91},
	year = {2019}}

@article{Collins:1981uw,
	author = {Collins, John C. and Soper, Davison E.},
	doi = {10.1016/0550-3213(82)90021-9},
	journal = {Nucl. Phys. B},
	pages = {445--492},
	reportnumber = {OITS-166},
	title = {{Parton Distribution and Decay Functions}},
	volume = {194},
	year = {1982}}

@article{Lee:2023xzv,
	archiveprefix = {arXiv},
	author = {Lee, Kyle and Moult, Ian and Ringer, Felix and Waalewijn, Wouter J.},
	eprint = {2308.00028},
	month = {7},
	primaryclass = {hep-ph},
	reportnumber = {MIT-CTP-5589, JLAB-THY-23-3891},
	title = {{A Formalism for Extracting Track Functions from Jet Measurements}},
	year = {2023}}

@article{LUCID2,
	author = {G. Avoni and others},
	doi = {10.1088/1748-0221/13/07/P07017},
	journal = {JINST},
	number = {07},
	pages = {P07017},
	title = {The new LUCID-2 detector for luminosity measurement and monitoring in ATLAS},
	volume = {13},
	year = {2018}}

@article{Webber:1983if,
	author = {Webber, B. R.},
	doi = {10.1016/0550-3213(84)90333-X},
	journal = {Nucl. Phys. B},
	pages = {492-528},
	reportnumber = {CERN-TH-3713},
	slaccitation = {%%CITATION = NUPHA,B238,492;%%},
	title = {{A QCD model for jet fragmentation including soft gluon interference}},
	volume = {238},
	year = {1984}}

@article{Bellm:2017bvx,
	archiveprefix = {arXiv},
	author = {Bellm, Johannes and others},
	eprint = {1705.06919},
	primaryclass = {hep-ph},
	reportnumber = {CERN-PH-TH-2017-109, CERN-TH-2017-109, MAN-HEP-2017-08, UWTHPH-2017-10, IFJPAN-IV-2017-7, NIKHEF-2017-026, HERWIG-2017-02, KA-TP-19-2017, MCNET-17-08, IPPP-17-40},
	slaccitation = {%%CITATION = ARXIV:1705.06919;%%},
	title = {{Herwig 7.1 Release Note}},
	year = {2017}}

@article{Malaescu:2009dm,
	archiveprefix = {arXiv},
	author = {Malaescu, Bogdan},
	eprint = {0907.3791},
	month = {7},
	primaryclass = {physics.data-an},
	reportnumber = {LAL-09-107},
	title = {{An iterative, dynamically stabilized method of data unfolding}},
	year = {2009}}

@article{jackknife,
	author = {B. Efron},
	issn = {00905364},
	journal = {The Annals of Statistics},
	number = {1},
	pages = {1--26},
	publisher = {Institute of Mathematical Statistics},
	title = {Bootstrap Methods: Another Look at the Jackknife},
	url = {http://www.jstor.org/stable/2958830},
	urldate = {2023-08-10},
	volume = {7},
	year = {1979}}

@article{CMS-SMP-22-007,
	archiveprefix = {arXiv},
	author = {{CMS Collaboration}},
	eprint = {2312.16343},
journal = {JHEP},
volume = {05},
pages = {116},
doi = {10.1007/JHEP05(2024)116},
	primaryclass = {hep-ex},
	reportnumber = {CERN-EP-2023-282},
	title = {{Measurement of the primary Lund jet plane density in proton--proton collisions at $\sqrt{s}$ = 13 TeV}},
	year = {2023}}

@article{STDM-2020-04,
	archiveprefix = {arXiv},
	 author = "{ATLAS Collaboration}",
journal = "Phys. Rev. D",
volume = "110",
year = "2024",
pages = "072019",
doi = "10.1103/PhysRevD.110.072019",
	eprint = {2405.20206},
	month = {5},
	primaryclass = {hep-ex},
	reportnumber = {CERN-EP-2024-119},
	title = {{Measurements of jet cross-section ratios in 13 TeV proton--proton collisions with ATLAS}},
	year = {2024}}

@Article{Cacciari:2008gp,
     author    = "Cacciari, Matteo and Salam, Gavin P. and Soyez, Gregory",
     title     = "{The anti-\(k_{t}\) jet clustering algorithm}",
     journal   = "JHEP",
     volume    = "04",
     year      = "2008",
     pages     = "063",
     eprint    = "0802.1189",
     archivePrefix = "arXiv",
     primaryClass  =  "hep-ph",
     doi       = "10.1088/1126-6708/2008/04/063",
     SLACcitation  = "%%CITATION = 0802.1189;%%"
}

@Article{Fastjet,
      author         = "Cacciari, Matteo and Salam, Gavin P. and Soyez, Gregory",
      title          = "{FastJet user manual}",
      journal        = "Eur. Phys. J. C",
      volume         = "72",
      year           = "2012",
      pages          = "1896",
      doi            = "10.1140/epjc/s10052-012-1896-2",
      eprint         = "1111.6097",
      archivePrefix  = "arXiv",
      primaryClass   = "hep-ph",
      reportNumber   = "CERN-PH-TH-2011-297",
      SLACcitation   = "%%CITATION = ARXIV:1111.6097;%%"
}

@article{Lai:2010vv,
      author         = "Lai, H.-L. and others",
      title          = "{New parton distributions for collider physics}",
      journal        = "Phys. Rev. D",
      volume         = "82",
      pages          = "074024",
      doi            = "10.1103/PhysRevD.82.074024",
      year           = "2010",
      eprint         = "1007.2241",
      archivePrefix  = "arXiv",
      primaryClass   = "hep-ph",
      reportNumber   = "MSUHEP-100707, SMU-HEP-10-10",
      SLACcitation   = "%%CITATION = ARXIV:1007.2241;%%",
}

@article{Dulat:2015mca,
      author         = "Dulat, Sayipjamal and Hou, Tie-Jiun and Gao, Jun and
                        Guzzi, Marco and Huston, Joey and Nadolsky, Pavel and
                        Pumplin, Jon and Schmidt, Carl and Stump, Daniel and Yuan,
                        C. P.",
      title          = "{New parton distribution functions from a global analysis
                        of quantum chromodynamics}",
      journal        = "Phys. Rev. D",
      volume         = "93",
      year           = "2016",
      number         = "3",
      pages          = "033006",
      doi            = "10.1103/PhysRevD.93.033006",
      eprint         = "1506.07443",
      archivePrefix  = "arXiv",
      primaryClass   = "hep-ph",
      SLACcitation   = "%%CITATION = ARXIV:1506.07443;%%"
}

@article{Harland-Lang:2014zoa,
      author         = "Harland-Lang, L. A. and Martin, A. D. and Motylinski, P.
                        and Thorne, R. S.",
      title          = "{Parton distributions in the LHC era: MMHT 2014 PDFs}",
      journal        = "Eur. Phys. J. C",
      number         = "5",
      volume         = "75",
      pages          = "204",
      doi            = "10.1140/epjc/s10052-015-3397-6",
      year           = "2015",
      eprint         = "1412.3989",
      archivePrefix  = "arXiv",
      primaryClass   = "hep-ph",
      reportNumber   = "LCTS-2014-47, IPPP-14-97, DCPT-14-194",
      SLACcitation   = "%%CITATION = ARXIV:1412.3989;%%",
}

@article{Ball:2012cx,
      author         = "{NNPDF Collaboration} and Ball, Richard D. and others",
      title          = "{Parton distributions with LHC data}",
      journal        = "Nucl. Phys. B",
      volume         = "867",
      year           = "2013",
      pages          = "244",
      doi            = "10.1016/j.nuclphysb.2012.10.003",
      eprint         = "1207.1303",
      archivePrefix  = "arXiv",
      primaryClass   = "hep-ph",
      reportNumber   = "EDINBURGH-2012-08, IFUM-FT-997, FR-PHENO-2012-014,
                        RWTH-TTK-12-25, CERN-PH-TH-2012-037, SFB-CPP-12-47\,
                        --CERN-PH-TH-2012-037",
      SLACcitation   = "%%CITATION = ARXIV:1207.1303;%%"
}

@Article{Sjostrand:2006za,
     author    = "Sj{\"o}strand, Torbjorn and Mrenna, Stephen and Skands, Peter",
     title     = "{PYTHIA 6.4 physics and manual}",
     journal   = "JHEP",
     volume    = "05",
     year      = "2006",
     pages     = "026",
     eprint    = "hep-ph/0603175",
     archivePrefix = "arXiv",
     doi       = "10.1088/1126-6708/2006/05/026",
     SLACcitation  = "%%CITATION = HEP-PH/0603175;%%"
}

@article{Sjostrand:2014zea,
      author         = "Sj{\"o}strand, Torbj{\"o}rn and Ask, Stefan and Christiansen,
                        Jesper R. and Corke, Richard and Desai, Nishita and Ilten,
                        Philip and Mrenna, Stephen and Prestel, Stefan and
                        Rasmussen, Christine O. and Skands, Peter Z.",
      title          = "{An introduction to PYTHIA 8.2}",
      journal        = "Comput. Phys. Commun.",
      volume         = "191",
      year           = "2015",
      pages          = "159",
      doi            = "10.1016/j.cpc.2015.01.024",
      eprint         = "1410.3012",
      archivePrefix  = "arXiv",
      primaryClass   = "hep-ph",
      reportNumber   = "LU-TP-14-36, MCNET-14-22, CERN-PH-TH-2014-190,
                        FERMILAB-PUB-14-316-CD, DESY-14-178, SLAC-PUB-16122,
                        --FERMILAB-PUB-14-316-CD",
      SLACcitation   = "%%CITATION = ARXIV:1410.3012;%%"
}

@article{Lonnblad:2001iq,
      author         = "L{\"o}nnblad, Leif",
      title          = "{Correcting the Colour-Dipole Cascade Model with Fixed Order Matrix Elements}",
      journal        = "JHEP",
      volume         = "05",
      year           = "2002",
      pages          = "046",
      doi            = "10.1088/1126-6708/2002/05/046",
      eprint         = "hep-ph/0112284",
      archivePrefix  = "arXiv",
      XprimaryClass   = "hep-ph",
      reportNumber   = "LU-TP-01-38",
      SLACcitation   = "%%CITATION = HEP-PH/0112284;%%"
}

@article{Lonnblad:2011xx,
      author         = "L{\"o}nnblad, Leif and Prestel, Stefan",
      title          = "{Matching tree-level matrix elements with interleaved
                        showers}",
      journal        = "JHEP",
      volume         = "03",
      year           = "2012",
      pages          = "019",
      doi            = "10.1007/JHEP03(2012)019",
      eprint         = "1109.4829",
      archivePrefix  = "arXiv",
      primaryClass   = "hep-ph",
      SLACcitation   = "%%CITATION = ARXIV:1109.4829;%%"
}

@Article{Lange:2001uf,
      author         = "Lange, D. J.",
      title          = "{The EvtGen particle decay simulation package}",
      booktitle      = "{Proceedings, 7th International Conference on B physics
                        at hadron machines (BEAUTY 2000)}",
      journal        = "Nucl. Instrum. Meth. A",
      volume         = "462",
      year           = "2001",
      pages          = "152",
      doi            = "10.1016/S0168-9002(01)00089-4",
      SLACcitation   = "%%CITATION = NUIMA,A462,152;%%"
}

@Article{Bahr:2008pv,
      author         = "B{\"a}hr, M. and others",
      title          = "{Herwig++ physics and manual}",
      journal        = "Eur. Phys. J. C",
      volume         = "58",
      year           = "2008",
      pages          = "639",
      doi            = "10.1140/epjc/s10052-008-0798-9",
      eprint         = "0803.0883",
      archivePrefix  = "arXiv",
      primaryClass   = "hep-ph",
      reportNumber   = "CERN-PH-TH-2008-038, CAVENDISH-HEP-08-03, KA-TP-05-2008,
                        DCPT-08-22, IPPP-08-11, CP3-08-05",
      SLACcitation   = "%%CITATION = ARXIV:0803.0883;%%"
}

@Article{Bellm:2015jjp,
      author         = "Bellm, Johannes and others",
      title          = "{Herwig 7.0/Herwig++ 3.0 release note}",
      journal        = "Eur. Phys. J. C",
      volume         = "76",
      year           = "2016",
      number         = "4",
      pages          = "196",
      doi            = "10.1140/epjc/s10052-016-4018-8",
      eprint         = "1512.01178",
      archivePrefix  = "arXiv",
      primaryClass   = "hep-ph",
      reportNumber   = "CERN-PH-TH-2015-289, MAN-HEP-2015-15, IFJPAN-IV-2015-13,
                        HERWIG-2015-01, KA-TP-18-2015, DCPT-15-142, MCNET-15-28,
                        IPPP-15-71, --HERWIG-2015-01",
      SLACcitation   = "%%CITATION = ARXIV:1512.01178;%%"
}

@Article{Gleisberg:2008ta,
      author         = "Gleisberg, T. and H{\"o}che, S. and Krauss, F. and Sch{\"o}nherr, M. and Schumann, S. and Siegert, F. and Winter J.",
      title          = "{Event generation with SHERPA 1.1}",
      journal        = "JHEP",
      volume         = "02",
      pages          = "007",
      doi            = "10.1088/1126-6708/2009/02/007",
      year           = "2009",
      eprint         = "0811.4622",
      archivePrefix  = "arXiv",
      primaryClass   = "hep-ph",
}

@Article{Schumann:2007mg,
      author         = "Schumann, Steffen and Krauss, Frank",
      title          = "{A parton shower algorithm based on Catani--Seymour dipole factorisation}",
      journal        = "JHEP",
      volume         = "03",
      pages          = "038",
      doi            = "10.1088/1126-6708/2008/03/038",
      year           = "2008",
      eprint         = "0709.1027",
      archivePrefix  = "arXiv",
      primaryClass   = "hep-ph",
}

@article{Agostinelli:2002hh,
      author         = "Agostinelli, S. and others",
      title          = "{\textsc{Geant4} -- a simulation toolkit}",
      journal        = "Nucl. Instrum. Meth. A",
      volume         = "506",
      year           = "2003",
      pages          = "250",
      doi            = "10.1016/S0168-9002(03)01368-8",
      reportNumber   = "SLAC-PUB-9350, FERMILAB-PUB-03-339",
      SLACcitation   = "%%CITATION = NUIMA,A506,250;%%"
}

@Inproceedings{Adye:2011gm,
    author = "Adye, Tim",
    title = "{Unfolding algorithms and tests using RooUnfold}",
    booktitle = "{Proceedings, 2011 Workshop on Statistical Issues Related to Discovery Claims in Search Experiments and Unfolding (PHYSTAT 2011)}",
    venue = "CERN, Geneva, Switzerland",
    eventdate = "2011-01-17/2011-01-20",
    pages = "313-318",
    eprint = "1105.1160",
    archivePrefix = "arXiv",
    primaryClass = "physics.data-an",
    doi = "10.5170/CERN-2011-006.313",
}

@article{DAgostini:1995,
  author = "Giulio D'Agostini",
  title = "A multidimensional unfolding method based on Bayes' theorem",
  journal = "Nucl. Instrum. Meth. A",
  volume = "362",
  number = "2",
  pages = "487 - 498",
  year = "1995",
  issn = "0168-9002",
  doi = "10.1016/0168-9002(95)00274-X",
}

@Article{PERF-2007-01,
    author         = "{ATLAS Collaboration}",
    title          = "{The ATLAS Experiment at the CERN Large Hadron Collider}",
    journal        = "JINST",
    volume         = "3",
    year           = "2008",
    pages          = "S08003",
    doi            = "10.1088/1748-0221/3/08/S08003",
    primaryClass   = "hep-ex",
}

@Article{SOFT-2010-01,
    author         = "{ATLAS Collaboration}",
    title          = "{The ATLAS Simulation Infrastructure}",
    journal        = "Eur. Phys. J. C",
    volume         = "70",
    year           = "2010",
    pages          = "823",
    doi            = "10.1140/epjc/s10052-010-1429-9",
    eprint         = "1005.4568",
    archivePrefix  = "arXiv",
    primaryClass   = "physics.ins-det",
}

@Article{STDM-2010-14,
    author         = "{ATLAS Collaboration}",
    title          = "{Properties of jets measured from tracks in proton--proton collisions at center-of-mass energy \(\sqrt{s} = 7\,\text{TeV}\) with the ATLAS detector}",
    journal        = "Phys. Rev. D",
    volume         = "84",
    year           = "2011",
    pages          = "054001",
    doi            = "10.1103/PhysRevD.84.054001",
    reportNumber   = "CERN-PH-EP-2011-110",
    eprint         = "1107.3311",
    archivePrefix  = "arXiv",
    primaryClass   = "hep-ex",
}

@Article{TRIG-2012-01,
    author         = "{ATLAS Collaboration}",
    title          = "{The performance of the jet trigger for the ATLAS detector during 2011 data taking}",
    journal        = "Eur. Phys. J. C",
    volume         = "76",
    year           = "2016",
    pages          = "526",
    doi            = "10.1140/epjc/s10052-016-4325-0",
    reportNumber   = "CERN-EP-2016-031",
    eprint         = "1606.07759",
    archivePrefix  = "arXiv",
    primaryClass   = "hep-ex",
}

@Article{STDM-2014-17,
    author         = "{ATLAS Collaboration}",
    title          = "{Measurement of jet charge in dijet events from \(\sqrt{s} = 8\,\text{TeV}\) \(pp\) collisions with the ATLAS detector}",
    journal        = "Phys. Rev. D",
    volume         = "93",
    year           = "2016",
    pages          = "052003",
    doi            = "10.1103/PhysRevD.93.052003",
    reportNumber   = "CERN-PH-EP-2015-207",
    eprint         = "1509.05190",
    archivePrefix  = "arXiv",
    primaryClass   = "hep-ex",
}

@Article{PERF-2015-07,
    author         = "{ATLAS Collaboration}",
    title          = "{Study of the material of the ATLAS inner detector for Run~2 of the LHC}",
    journal        = "JINST",
    volume         = "12",
    year           = "2017",
    pages          = "P12009",
    doi            = "10.1088/1748-0221/12/12/P12009",
    reportNumber   = "CERN-EP-2017-081",
    eprint         = "1707.02826",
    archivePrefix  = "arXiv",
    primaryClass   = "hep-ex",
}

@Article{PERF-2015-08,
    author         = "{ATLAS Collaboration}",
    title          = "{Performance of the ATLAS track reconstruction algorithms in dense environments in LHC Run~2}",
    journal        = "Eur. Phys. J. C",
    volume         = "77",
    year           = "2017",
    pages          = "673",
    doi            = "10.1140/epjc/s10052-017-5225-7",
    reportNumber   = "CERN-EP-2017-045",
    eprint         = "1704.07983",
    archivePrefix  = "arXiv",
    primaryClass   = "hep-ex",
}

@Article{PERF-2015-09,
    author         = "{ATLAS Collaboration}",
    title          = "{Jet reconstruction and performance using particle flow with the ATLAS Detector}",
    journal        = "Eur. Phys. J. C",
    volume         = "77",
    year           = "2017",
    pages          = "466",
    doi            = "10.1140/epjc/s10052-017-5031-2",
    reportNumber   = "CERN-EP-2017-024",
    eprint         = "1703.10485",
    archivePrefix  = "arXiv",
    primaryClass   = "hep-ex",
}

@Article{TRIG-2016-01,
    author         = "{ATLAS Collaboration}",
    title          = "{Performance of the ATLAS trigger system in 2015}",
    journal        = "Eur. Phys. J. C",
    volume         = "77",
    year           = "2017",
    pages          = "317",
    doi            = "10.1140/epjc/s10052-017-4852-3",
    reportNumber   = "CERN-EP-2016-241",
    eprint         = "1611.09661",
    archivePrefix  = "arXiv",
    primaryClass   = "hep-ex",
}

@Article{STDM-2017-16,
    author         = "{ATLAS Collaboration}",
    title          = "{Properties of jet fragmentation using charged particles measured with the ATLAS detector in \(pp\) collisions at \(\sqrt{s} = 13\,\text{TeV}\)}",
    journal        = "Phys. Rev. D",
    volume         = "100",
    year           = "2019",
    pages          = "052011",
    doi            = "10.1103/PhysRevD.100.052011",
    reportNumber   = "CERN-EP-2019-090",
    eprint         = "1906.09254",
    archivePrefix  = "arXiv",
    primaryClass   = "hep-ex",
}

@Article{STDM-2017-33,
    author         = "{ATLAS Collaboration}",
    title          = "{Measurement of soft-drop jet observables in \(pp\) collisions with the ATLAS detector at \(\sqrt{s} = 13\,\text{TeV}\)}",
    journal        = "Phys. Rev. D",
    volume         = "101",
    year           = "2020",
    pages          = "052007",
    doi            = "10.1103/PhysRevD.101.052007",
    reportNumber   = "CERN-EP-2019-269",
    eprint         = "1912.09837",
    archivePrefix  = "arXiv",
    primaryClass   = "hep-ex",
}

@Article{JETM-2018-02,
    author         = "{ATLAS Collaboration}",
    title          = "{In situ calibration of large-radius jet energy and mass in \(13\,\text{TeV}\) proton--proton collisions with the ATLAS detector}",
    journal        = "Eur. Phys. J. C",
    volume         = "79",
    year           = "2019",
    pages          = "135",
    doi            = "10.1140/epjc/s10052-019-6632-8",
    reportNumber   = "CERN-EP-2018-191",
    eprint         = "1807.09477",
    archivePrefix  = "arXiv",
    primaryClass   = "hep-ex",
}

@Article{JETM-2018-05,
    author         = "{ATLAS Collaboration}",
    title          = "{Jet energy scale and resolution measured in proton--proton collisions at \(\sqrt{s} = 13\,\text{TeV}\) with the ATLAS detector}",
    journal        = "Eur. Phys. J. C",
    volume         = "81",
    year           = "2021",
    pages          = "689",
    doi            = "10.1140/epjc/s10052-021-09402-3",
    reportNumber   = "CERN-EP-2020-083",
    eprint         = "2007.02645",
    archivePrefix  = "arXiv",
    primaryClass   = "hep-ex",
}

@Article{STDM-2018-57,
    author         = "{ATLAS Collaboration}",
    title          = "{Measurement of the Lund Jet Plane Using Charged Particles in \(13\,\text{TeV}\) Proton--Proton Collisions with the ATLAS Detector}",
    journal        = "Phys. Rev. Lett.",
    volume         = "124",
    year           = "2020",
    pages          = "222002",
    doi            = "10.1103/PhysRevLett.124.222002",
    reportNumber   = "CERN-EP-2020-030",
    eprint         = "2004.03540",
    archivePrefix  = "arXiv",
    primaryClass   = "hep-ex",
}

@Booklet{ATL-SOFT-PUB-2025-001,
    author         = "{ATLAS Collaboration}",
    title          = "{ATLAS Computing Acknowledgements}",
    howpublished   = "{ATL-SOFT-PUB-2025-001}",
    url            = "https://cds.cern.ch/record/2922210",
    year           = 2025
}

@Article{CMS-SMP-20-010,
    author         = "{CMS Collaboration}",
    title          = "{Study of quark and gluon jet substructure in \(Z\)+jet and dijet events from \(pp\) collisions}",
    journal        = "JHEP",
    volume         = "01",
    year           = "2022",
    pages          = "188",
    doi            = "10.1007/JHEP01(2022)188",
    reportNumber   = "CERN-EP-2021-161",
    eprint         = "2109.03340",
    archivePrefix  = "arXiv",
    primaryClass   = "hep-ex",
}

@Booklet{ATLAS-CONF-2015-029,
    author         = "{ATLAS Collaboration}",
    title          = "{Selection of jets produced in \(13~\text{TeV}\) proton--proton collisions with the ATLAS detector}",
    howpublished   = "{ATLAS-CONF-2015-029}",
    url            = "https://cds.cern.ch/record/2037702",
    year           = "2015",
}

@Booklet{ATL-PHYS-PUB-2014-021,
    author         = "{ATLAS Collaboration}",
    title          = "{ATLAS Pythia~8 tunes to \(7~\text{TeV}\) data}",
    howpublished   = "{ATL-PHYS-PUB-2014-021}",
    url            = "https://cds.cern.ch/record/1966419",
    year           = "2014",
}

@Booklet{ATL-PHYS-PUB-2015-006,
    author         = "{ATLAS Collaboration}",
    title          = "{The Optimization of ATLAS Track Reconstruction in Dense Environments}",
    howpublished   = "{ATL-PHYS-PUB-2015-006}",
    url            = "https://cds.cern.ch/record/2002609",
    year           = "2015",
}

@Booklet{ATL-PHYS-PUB-2015-026,
    author         = "{ATLAS Collaboration}",
    title          = "{Vertex Reconstruction Performance of the ATLAS Detector at \(\sqrt{s} = 13~\text{TeV}\)}",
    howpublished   = "{ATL-PHYS-PUB-2015-026}",
    url            = "https://cds.cern.ch/record/2037717",
    year           = "2015",
}

@Booklet{ATL-PHYS-PUB-2015-044,
    author         = "{ATLAS Collaboration}",
    title          = "{Measurement of performance of the pixel neural network clustering algorithm of the ATLAS experiment at \(\sqrt{s} = 13~\text{TeV}\)}",
    howpublished   = "{ATL-PHYS-PUB-2015-044}",
    url            = "https://cds.cern.ch/record/2054921",
    year           = "2015",
}

@Booklet{ATL-PHYS-PUB-2015-052,
    author         = "{ATLAS Collaboration}",
    title          = "{Robustness of the Artificial Neural Network Clustering Algorithm of the ATLAS experiment}",
    howpublished   = "{ATL-PHYS-PUB-2015-052}",
    url            = "https://cds.cern.ch/record/2116350",
    year           = "2015",
}

@Booklet{ATL-PHYS-PUB-2016-017,
    author         = "{ATLAS Collaboration}",
    title          = "{The Pythia~8 A3 tune description of ATLAS minimum bias and inelastic measurements incorporating the Donnachie--Landshoff diffractive model}",
    howpublished   = "{ATL-PHYS-PUB-2016-017}",
    url            = "https://cds.cern.ch/record/2206965",
    year           = "2016",
}

@Booklet{ATL-PHYS-PUB-2017-016,
    author         = "{ATLAS Collaboration}",
    title          = "{Modelling of Track Reconstruction Inside Jets with the 2016 ATLAS \(\sqrt{s} = 13~\text{TeV}\) \(pp\) Dataset}",
    howpublished   = "{ATL-PHYS-PUB-2017-016}",
    url            = "https://cds.cern.ch/record/2275639",
    year           = "2017",
}

@Booklet{ATL-PHYS-PUB-2019-017,
    author         = "{ATLAS Collaboration}",
    title          = "{Multijet simulation for \(13\,\text{TeV}\) ATLAS Analyses}",
    howpublished   = "{ATL-PHYS-PUB-2019-017}",
    url            = "https://cds.cern.ch/record/2672252",
    year           = "2019",
}
\clearpage
 
\begin{flushleft}
\hypersetup{urlcolor=black}
{\Large The ATLAS Collaboration}

\bigskip

\AtlasOrcid[0000-0002-6665-4934]{G.~Aad}$^\textrm{\scriptsize 104}$,
\AtlasOrcid[0000-0001-7616-1554]{E.~Aakvaag}$^\textrm{\scriptsize 17}$,
\AtlasOrcid[0000-0002-5888-2734]{B.~Abbott}$^\textrm{\scriptsize 123}$,
\AtlasOrcid[0000-0002-0287-5869]{S.~Abdelhameed}$^\textrm{\scriptsize 119a}$,
\AtlasOrcid[0000-0002-1002-1652]{K.~Abeling}$^\textrm{\scriptsize 56}$,
\AtlasOrcid[0000-0001-5763-2760]{N.J.~Abicht}$^\textrm{\scriptsize 50}$,
\AtlasOrcid[0000-0002-8496-9294]{S.H.~Abidi}$^\textrm{\scriptsize 30}$,
\AtlasOrcid[0009-0003-6578-220X]{M.~Aboelela}$^\textrm{\scriptsize 45}$,
\AtlasOrcid[0000-0002-9987-2292]{A.~Aboulhorma}$^\textrm{\scriptsize 36e}$,
\AtlasOrcid[0000-0001-5329-6640]{H.~Abramowicz}$^\textrm{\scriptsize 155}$
% ...
\bigskip
\\

% ...
$^{17}$Department for Physics and Technology, University of Bergen, Bergen; Norway.\\
% ...
$^{30}$Physics Department, Brookhaven National Laboratory, Upton NY; United States of America.\\
% ...
$^{36}$$^{(a)}$Facult\'e des Sciences Ain Chock, Universit\'e Hassan II de Casablanca;$^{(b)}$Facult\'{e} des Sciences, Universit\'{e} Ibn-Tofail, K\'{e}nitra;$^{(c)}$Facult\'e des Sciences Semlalia, Universit\'e Cadi Ayyad, LPHEA-Marrakech;$^{(d)}$LPMR, Facult\'e des Sciences, Universit\'e Mohamed Premier, Oujda;$^{(e)}$Facult\'e des sciences, Universit\'e Mohammed V, Rabat;$^{(f)}$Institute of Applied Physics, Mohammed VI Polytechnic University, Ben Guerir; Morocco.\\
% ...
$^{45}$Physics Department, Southern Methodist University, Dallas TX; United States of America.\\
% ...
$^{50}$Fakult\"{a}t Physik , Technische Universit{\"a}t Dortmund, Dortmund; Germany.\\
% ...
$^{56}$II. Physikalisches Institut, Georg-August-Universit\"{a}t G\"ottingen, G\"ottingen; Germany.\\
% ...
$^{104}$CPPM, Aix-Marseille Universit\'e, CNRS/IN2P3, Marseille; France.\\
% ...
$^{119}$$^{(a)}$New York University Abu Dhabi, Abu Dhabi;$^{(b)}$United Arab Emirates University, Al Ain; United Arab Emirates.\\
% ...
$^{123}$Homer L. Dodge Department of Physics and Astronomy, University of Oklahoma, Norman OK; United States of America.\\
% ...
$^{155}$Raymond and Beverly Sackler School of Physics and Astronomy, Tel Aviv University, Tel Aviv; Israel.\\
% ...
$^{a}$ Also Affiliated with an institute covered by a cooperation agreement with CERN.\\
$^{b}$ Also at An-Najah National University, Nablus; Palestine.\\
$^{c}$ Also at Borough of Manhattan Community College, City University of New York, New York NY; United States of America.\\
$^{d}$ Also at Center for High Energy Physics, Peking University; China.\\
$^{e}$ Also at Center for Interdisciplinary Research and Innovation (CIRI-AUTH), Thessaloniki; Greece.\\
$^{f}$ Also at CERN, Geneva; Switzerland.\\
% ...

\end{flushleft}


%

%
%

%

\end{document}